\documentclass[10pt]{article}
\usepackage{bm,latexsym,amsmath,amsthm,amssymb}
\usepackage{graphicx,indentfirst}
\usepackage[numbers,sort&compress]{natbib}
\date{}
\begin{document}
\title{\bf Existence and uniqueness of global weak solutions to a generalized Camassa-Holm equation}
\author{
Qiaoling Chen\\
\small School of Mathematics and Information Science, Shaanxi Normal University,
Xi'an 710062, PR China\\
\small School of Science, Xi'an Polytechnic University,
Xi'an 710048, PR China\\[3pt]
Feng Wang\thanks{Corresponding author\newline
\mbox{}\qquad E-mail: wangfeng@xidian.edu.cn}\\
\small School of Mathematics and Statistics, Xidian University,
    Xi'an 710071, PR China}
\date{}
\maketitle \baselineskip 3pt
\begin{center}
\begin{minipage}{130mm}
{{\bf Abstract.} This paper is concerned with the existence and uniqueness of global weak solutions to a generalized Camassa-Holm equation on real line. By introducing some new variables, the equation is transformed into two different semi-linear systems. Then the existence and uniqueness of global weak solutions to the original equation are obtained from that of the two semi-linear systems, respectively.

\vskip 0.2cm{\bf Keywords:} Generalized Camassa-Holm equation; Global weak solution; Characteristcs; Existence; Uniqueness.}

\vskip 0.2cm{\bf AMS subject classifications (2000):} 35L05, 35D30.
\end{minipage}
\end{center}

\baselineskip=15pt

\section{Introduction}
\label{intro}

In recent years, the shallow-water wave equations have attracted much attention. The failure of weakly nonlinear shallow-water wave equations, such as the well-known Kortewegde Vries (KdV) and Boussinesq equations, to model some interesting physical phenomena like wave breaking and high-amplitude waves is prime motivation for transition to full nonlinearity in the search for alternative models for nonlinear shallow-water waves. With the aid of an asymptotic approximation to the Hamiltonian of the Green-Naghdi (GN) equations, Camassa and Holm in 1993 derived the following Camassa-Holm (CH) equation \cite{ch93}
$$
\begin{array}{l}
u_{t}-u_{txx}+3uu_{x}-2u_{x}u_{xx}-uu_{xxx}=0,
\end{array}
\eqno(1.1)
$$
which has both solitary waves interacting like solitons and, in contrast to
KdV, solutions which blow up in finite time as a result of the breaking
of waves. Equation (1.1) was actually obtained much earlier as an abstract bi-Hamiltonian equation with infinitely many conservation laws by Fokas and Fuchssteiner \cite{ff81}, and was also found independently by Dai \cite{dai98} as a model for nonlinear waves in cylindrical hyperelastic rods. From the viewpoint of geometry, it is a re-expression of the geodesic flow both on the diffeomorphism group of the circle \cite{ck02} and on the Bott-Virasoro group \cite{k07}. Moreover, it has been extended to an entire integrable hierarchy including both negative and positive flows and shown to admit algebro-geometric solutions on a symplectic submanifold \cite{Qiao2003}.

It is convenient to equivalently write the CH equation (1.1) in the following
nonlocal form
$$
\begin{array}{l}
u_{t}+uu_{x}+\partial_{x}p*(u^{2}+\frac{1}{2}u_{x}^{2})=0,
\end{array}
\eqno(1.2)
$$
where $p(x)=\frac{1}{2}e^{-|x|}$ with $x\in \mathbb{R}$ satisfies $p*f=(1-\partial_{x}^{2})^{-1}f$ for all $f\in L^{2}(\mathbb{R})$.
The Cauchy problem of (1.2), in particular its well-posedness, blow-up behavior and global existence, have been well-studied both on the real line and on the circle, e.g., \cite{brco07,bcz15,brc07,c97,ce98,cm99,d01,cs00,ca00,cm00,esy08,hora07,hora08,lo00,xz00,ces98,ce00,himk09,himkm10}.
Equation (1.2) with weakly dissipative term, which is of the form
$$
\begin{array}{l}
u_{t}+uu_{x}+\partial_{x}p*(u^{2}+\frac{1}{2}u_{x}^{2})+\lambda u=0,\quad
\lambda>0,
\end{array}
\eqno(1.3)
$$
was studied in \cite{wy09}. Unlike (1.2), equation (1.3) has no traveling wave solution and its $H^{1}$-energy is not conserved. However, they all possess global solutions and have the same blow-up rate.
Moreover, equation (1.2) with a forcing, which is of the form
$$
\begin{array}{l}
u_{t}+uu_{x}+\partial_{x}p*(u^{2}+\frac{1}{2}u_{x}^{2})+kp*u=0,\quad
k\in \mathbb{R},
\end{array}
\eqno(1.4)
$$
was proved to admits unique global weak solution in \cite{z16}.

The CH equation was also derived by Constantin and Lannes in \cite{cola09} as asymptotical equation to the GN equations under the Camassa-Holm scaling. When the effect of solid-body rotation of the Earth, namely the Coriolis effect, is considered, by following the idea of \cite{cola09}, Chen et al. \cite{cgl18} recently derived the following rotation-Camassa-Holm (R-CH) equation as asymptotical equation to the rotation-Green-Naghdi (R-GN) equations with the Coriolis effect under the Camassa-Holm scaling
$$
\begin{array}{l}
u_{t}+(u+\frac{\beta_{0}}{\beta})u_{x}
+\partial_{x}p*\left((c-\frac{\beta_{0}}{\beta})u+u^{2}
+\frac{\omega_{1}}{3\alpha^{2}}u^{3}+\frac{\omega_{2}}{4\alpha^{3}}u^{4}+\frac{1}{2}u_{x}^{2}
\right)=0,
\end{array}
\eqno(1.5)
$$
which has a cubic and even quartic nonlinearities and a formal Hamiltonian structure. Equation (1.5) was also derived in \cite{gls19} as a model equation which describes the motion of the fluid with the Coriolis effect from the incompressible shallow water in the equatorial region.

In this paper, we consider the following generalized Camassa-Holm equation
$$
\begin{array}{l}
u_{t}+(\alpha u+\beta)u_{x}+\partial_{x}p*\left(h(u)+\frac{\alpha}{2}u_{x}^{2}\right)
+kp*u+\lambda u=0,\quad x\in \mathbb{R},
\end{array}
\eqno(1.6)
$$
where $\alpha,\beta, k,\lambda\in \mathbb{R}$ are constants and $h: \mathbb{R}\rightarrow \mathbb{R}$ is a given locally Lipschitz function with $h(0)=0$. Motivated by the works on the CH equation (1.2) in \cite{brco07,bcz15}, equation (1.4) in \cite{z16} and the Novikov equation in \cite{ccl18}, we aim to investigate the issue on the existence and uniqueness of global weak solutions to (1.6). Equation (1.6) is more general than (1.2)-(1.5) and can be equivalently rewritten as
$$
\begin{array}{l}
u_{t}+(\frac{\alpha}{2} u^{2}+\beta u)_{x}+(\lambda+k)u
+\partial_{x}p*\left(h(u)+ku_{x}+\frac{\alpha}{2}u_{x}^{2}\right)
=0,
\end{array}
$$
which is a special inviscid case of the model studied by Coclite et al. in \cite{chka05} where the global existence and uniqueness of smooth solutions
were proved. When $\beta=k=\lambda=0$ in (1.6), the Peakons in a particular case were studied in \cite{qt01} and their stability was discussed in \cite{lop03}, the precise blow-up scenario was established by Yin \cite{yi04}, the existence of a strongly continuous semigroup of global weak solutions was investigated by Coclite et al. \cite{chkar06}, and the existence and uniqueness of global conservative weak solutions were showed by Zhou et al.\cite{zm13,yrzm18}. However, in our model (1.6) the forcing terms $kp*u$ and $\lambda u$ destroy the conservation of $H^{1}$-energy, so we can only consider global weak solutions that are not conservative.

One of the main difficulties in model (1.6) is that the term $\lambda u$ has a significant impact on the balance law (see (1.9) later) which plays a key role in the uniqueness, this suggests us to define new Radon measures $\mu_{(t)}$ whose absolutely continuous part w.r.t. Lebesgue measure have density $e^{2\lambda t}u_{x}^{2}(t,\cdot)$, rather than $u_{x}^{2}(t,\cdot)$ used in the previous works \cite{brco07,bcz15,tlm19,yrzm18,zm13,mo07,zm14}. Another difficulty is that $h(u)$ and $P_{2}=kp*u$ may contain linear term $u$, which requires us to make finer estimates for some terms according to the $H^{1}$-energy of $u$ in proving the global existence of solutions to semi-linear system, e.g., see $\|P_{1}(T)\|_{L^{\infty}}$ in (3.10) and $\|\partial_{x}P_{2}(T)\|_{L^{\infty}}$ in (3.11) later. Finally, it is worth mentioning that the weakly dissipative CH equation (1.3) ($\lambda>0$) has been showed to admit global weak solutions by compactness methods in \cite{wu11}, but in our paper we assume $\lambda\in \mathbb{R}$ and obtain the uniqueness results.

Now we state our main results for the existence and uniqueness of global weak solutions to (1.6). We define
$$
P_{1}=p*\left(h(u)+\frac{\alpha}{2}u_{x}^{2}\right),\quad
P_{2}=kp*u,
$$
then the initial value problem of (1.6) becomes into
$$
\begin{array}{l}
\left\{\begin{array}{l}
u_{t}+(\alpha u+\beta)u_{x}+\partial_{x}P_{1}+P_{2}+\lambda u=0,\\[3pt]
u(0,x)=u_{0}(x).
\end{array}
\right.
\end{array}
\eqno(1.7)
$$
For smooth solutions, we differentiate the equation in (1.7) with respect to $x$ to get
$$
\begin{array}{l}
u_{tx}+(\alpha u+\beta)u_{xx}+\frac{\alpha}{2}u_{x}^{2}
-h(u)+P_{1}+\partial_{x}P_{2}+\lambda u_{x}=0.
\end{array}
\eqno(1.8)
$$
Multiplying $u_{x}$ to (1.8), we have
$$
(u_{x}^{2})_{t}+((\alpha u+\beta)u_{x}^{2})_{x}
+2(-h(u)+P_{1}+\partial_{x}P_{2})u_{x}+2\lambda u_{x}^{2}=0,
$$
or equivalent form
$$
\begin{array}{l}
(e^{2\lambda t}u_{x}^{2})_{t}+e^{2\lambda t}((\alpha u+\beta)u_{x}^{2})_{x}
+2e^{2\lambda t}(-h(u)+P_{1}+\partial_{x}P_{2})u_{x}=0,
\end{array}
\eqno(1.9)
$$
which is called the balance law. \\

\noindent\textbf{Theorem 1.1.}  Let $u_{0}\in H^{1}(\mathbb{R})$ be an absolutely continuous function on $x$. Then the Cauchy problem (1.7) admits a global weak solution $u(t,x)\in H^{1}(\mathbb{R})$ satisfying the initial data  in $L^{2}(\mathbb{R})$ together with
$$
\begin{array}{l}
\int_{\Gamma}\left(-u_{x}\phi_{t}-(\alpha u+\beta)u_{x}\phi_{x}
+(P_{1}+\partial_{x}P_{2}-h(u)-\frac{\alpha}{2}u_{x}^{2}+\lambda u_{x})\phi\right)dxdt
-\int_{\mathbb{R}}u_{0,x}\phi(0,x)dx=0
\end{array}
\eqno(1.10)
$$
for every text function $\phi\in C_{c}^{1}(\Gamma)$ with $\Gamma=\{(t,x); t\geq 0, x\in \mathbb{R}\}$. Moreover, the weak solution satisfies the following properties:

$(i)$ on any bounded time interval $u(t,x)$ is H\"{o}lder continuous with exponent $\frac{1}{2}$ w.r.t. $t$ and $x$;

$(ii)$ the map $t\mapsto u(t,\cdot)$ is Lipschitz continuous under $L^{2}$-norm;

$(iii)$ the balance law (1.9) is satisfied in the following sense: there exists a family of Randon measures $\{\mu_{(t)}, t\in \mathbb{R}\}$, depending continuously on time w.r.t. the topology of weak convergence of measures, and for every $t\in \mathbb{R}$, the absolutely continuous part of $\mu_{(t)}$ w.r.t. Lebesgue measure has density $e^{2\lambda t}u_{x}^{2}(t,\cdot)$, which provides a measure-valued solution to the balance law
$$
\begin{array}{l}
\int_{\mathbb{R}^{+}}\left(\int[\phi_{t}+(\alpha u+\beta)\phi_{x}]d\mu_{(t)}
+\int 2e^{2\lambda t}(h(u)-P_{1}-\partial_{x}P_{2})u_{x}\phi dx\right)dt
+\int_{\mathbb{R}}u_{0,x}^{2}\phi(0,x)dx=0
\end{array}
\eqno(1.11)
$$
for every test function $\phi\in C_{c}^{1}(\Gamma)$.

$(iv)$ the solution depend continuously on the initial data. That is, for a sequence of initial data $u_{0,n}$ such that $\|u_{0,n}-u_{0}\|_{H^{1}}\rightarrow 0$ as $n\rightarrow \infty$, the corresponding solution $u_{n}(t,x)$ converge to $u(t,x)$ in any bounded sets.\\

\noindent\textbf{Theorem 1.2.} Let $u_{0}\in H^{1}(\mathbb{R})$ be an absolutely continuous function on $x$. Then the Cauchy problem (1.7) admits a unique global weak solution satisfying the initial data in $L^{2}(\mathbb{R})$ together with (1.10) and (1.11).\\

We remark that the approachs in \cite{brco07,bcz15} have also been used to prove the existence and uniqueness of global conservative weak solutions to two-component CH (CH2) system \cite{lz17,whc10} and modified two-component CH (MCH2) system \cite{ct18,g18,ty11}. Moreover, Holden et al. \cite{ghr12,hora07,hora08} reformulated the CH equation and CH2 system to semilinear systems of ordinary differential equations by means of the transformation between Eulerian and Lagrangian coordinates, and obtained the global existence of conservative weak solutions both on the real line and on the circle.

The rest of the paper is organized as follows. In Section 2, along the characteristic, we transfer the equation (1.6) to a semi-linear system by introducing some new variables. In Section 3, we first prove the local existence of solutions to the semi-linear system by applying the standard ODE
theory and extend it to the global one. Then we transform the solution for the semi-linear system to the original problem (1.7). Uniqueness of the global weak solution is established in Section 4.

\section{Semi-linear system for smooth solutions}

In this section, we derive a semi-linear system for smooth solutions by introducing some new variables.

The equation of the characteristic is
$$
\frac{dx(t)}{dt}=\alpha u(t,x(t))+\beta.
$$
We denote the characteristic passing through the point $(t,x)$ as
$$
s\mapsto x^{c}(s; t,x),
$$
and use the energy density $1+u_{x}^{2}$ to define the characteristic coordinate $Y$
$$
\begin{array}{rl}
Y\equiv Y(t,x):
=\int_{0}^{x^{c}(0; t,x)}(1+u_{x}^{2}(0, \bar{x}))d\bar{x}.
\end{array}
\eqno(2.1)
$$
Since
$$
x^{c}(0; t,x)=x-\int_{0}^{t}(\alpha u(s,x(s))+\beta)ds,
$$
it follows from (2.1) that
$$
\begin{array}{rl}
Y_{t}+(\alpha u+\beta)Y_{x}=0, \quad \forall~(t,x)\in \mathbb{R}^{+}\times\mathbb{R}.
\end{array}
\eqno(2.2)
$$

We also define $T=t$ to obtain the new coordinate $(T, Y)$. Thus, any smooth function
$f(t, x)=f(T, x(T,Y))$ can be considered as a function of $(T,Y)$ and also denoted by $f(T,Y)$. It is easy to check that
$$
\begin{array}{rl}
f_{t}+(\alpha u+\beta)f_{x}
&=f_{Y}(Y_{t}+(\alpha u+\beta)Y_{x})
+f_{T}(T_{t}+(\alpha u+\beta)T_{x})
=f_{T},\\[3pt]
f_{x}
&=f_{Y}Y_{x}+f_{T}T_{x}
=f_{Y}Y_{x}.
\end{array}
\eqno(2.3)
$$

We define new variables $v=v(T,Y)$ and $\xi=\xi(T,Y)$ as follows
$$
\begin{array}{rl}
v:=2\arctan u_{x}, \quad \xi:=\frac{1+u_{x}^{2}}{Y_{x}},
\end{array}
\eqno(2.4)
$$
where $u_{x}=u_{x}(T,x(T,Y))$. Simple computation yields
$$
\frac{1}{1+u_{x}^{2}}=\cos^{2}\frac{v}{2},\quad
\frac{u_{x}^{2}}{1+u_{x}^{2}}=\sin^{2}\frac{v}{2},\quad
\frac{u_{x}}{1+u_{x}^{2}}=\frac{1}{2}\sin v,\quad
x_{Y}=\frac{1}{Y_{x}}=\xi\cos^{2}\frac{v}{2}.
$$
Then, we will consider (1.7) under the new characteristic coordinate $(T,Y)$.
First, by (2.3), we have
$$
u_{T}
=u_{t}+(\alpha u+\beta)u_{x}
=-\partial_{x}P_{1}-P_{2}-\lambda u
$$
with
$$
\begin{array}{rl}
P_{1}(T,Y)
&=\frac{1}{2}\int_{-\infty}^{+\infty}
e^{-|x(T,Y)-\bar{x}|}(h(u)+\frac{\alpha}{2}u_{x}^{2})(t,\bar{x})d\bar{x}\\[5pt]
&=\frac{1}{2}\int_{-\infty}^{+\infty}
e^{-|\int_{\bar{Y}}^{Y}(\xi\cos^{2}\frac{v}{2})(T,\hat{Y})d\hat{Y}|}
(\xi h(u)\cos^{2}\frac{v}{2}
+\frac{\alpha}{2}\xi\sin^{2}\frac{v}{2})(T,\bar{Y})d\bar{Y},\\[5pt]
\partial_{x}P_{1}(T,Y)
&=\frac{1}{2}(\int_{Y}^{+\infty}-\int_{-\infty}^{Y})
e^{-|\int_{\bar{Y}}^{Y}(\xi\cos^{2}\frac{v}{2})(T,\hat{Y})d\hat{Y}|}
(\xi h(u)\cos^{2}\frac{v}{2}+\frac{\alpha}{2}\xi\sin^{2}\frac{v}{2})(T,\bar{Y})d\bar{Y},\\[5pt]
P_{2}(T,Y)
&=\frac{k}{2}\int_{-\infty}^{+\infty}
e^{-|\int_{\bar{Y}}^{Y}(\xi\cos^{2}\frac{v}{2})(T,\hat{Y})d\hat{Y}|}
(\xi u\cos^{2}\frac{v}{2})(T,\bar{Y})d\bar{Y},\\[5pt]
\partial_{x}P_{2}(T,Y)
&=\frac{k}{2}(\int_{Y}^{+\infty}-\int_{-\infty}^{Y})
e^{-|\int_{\bar{Y}}^{Y}(\xi\cos^{2}\frac{v}{2})(T,\hat{Y})d\hat{Y}|}
(\xi u\cos^{2}\frac{v}{2})(T,\bar{Y})d\bar{Y}.
\end{array}
\eqno(2.5)
$$
From (2.3)-(2.4), we can deduce that
$$
\begin{array}{rl}
v_{T}
&=\frac{2}{1+u_{x}^{2}}(u_{x})_{T}
=\frac{2}{1+u_{x}^{2}}(u_{xt}+(\alpha u+\beta)u_{xx})
=\frac{2}{1+u_{x}^{2}}(-\frac{\alpha}{2}u_{x}^{2}
+h(u)-P_{1}-\partial_{x}P_{2}-\lambda u_{x})\\[3pt]
&=-\alpha\sin^{2}\frac{v}{2}
+2h(u)\cos^{2}\frac{v}{2}
-2(P_{1}+\partial_{x}P_{2})\cos^{2}\frac{v}{2}-\lambda \sin v.
\end{array}
$$

Next, we derive the equation for $\xi$ by using the following relation
$$
Y_{tx}+(\alpha u+\beta)Y_{xx}=-\alpha u_{x}Y_{x},
$$
which can be deduced from (2.2). By (2.3)-(2.4), we have
$$
\begin{array}{rl}
\xi_{T}
&=\frac{2u_{x}}{Y_{x}}(u_{tx}+(\alpha u+\beta)u_{xx})
+\frac{-(1+u_{x}^{2})}{Y_{x}^{2}}(Y_{tx}+(\alpha u+\beta)Y_{xx})\\[3pt]
&=\frac{2u_{x}(u_{tx}+(\alpha u+\beta)u_{xx}+\frac{\alpha}{2}(1+u_{x}^{2}))}{Y_{x}}
=\frac{2u_{x}(\frac{\alpha}{2}
+h(u)-P_{1}-\partial_{x}P_{2}-\lambda u_{x}))}
{Y_{x}}\\[3pt]
&=\xi(\frac{\alpha}{2}+h(u)-P_{1}-\partial_{x}P_{2})\sin v
-2\lambda\xi\sin^{2}\frac{v}{2}.
\end{array}
$$
In conclusion, we transfer the quasi-linear equation (1.7) to the following semi-linear system on unknown variables $u,v$ and $\xi$ under the new coordinate $(T,Y)$:
$$
\begin{array}{l}
\left\{\begin{array}{l}
u_{T}=-\partial_{x}P_{1}-P_{2}-\lambda u, \\[3pt]
v_{T}=-\alpha\sin^{2}\frac{v}{2}
+2h(u)\cos^{2}\frac{v}{2}
-2(P_{1}+\partial_{x}P_{2})\cos^{2}\frac{v}{2}-\lambda \sin v,\\[3pt]
\xi_{T}=\xi(\frac{\alpha}{2}+h(u)-P_{1}-\partial_{x}P_{2})\sin v
-2\lambda\xi\sin^{2}\frac{v}{2}.
\end{array}
\right.
\end{array}
\eqno(2.6)
$$

\section{Global existence}

In this section, we first prove the global existence of solutions for the semi-linear system (2.6), and then transform the solution for (2.6) to the original problem (1.7).

\subsection{Global existence of semi-linear system}

In this subsection, we study the global existence of solutions to the following semi-linear system derived in the previous section
$$
\begin{array}{l}
\left\{\begin{array}{l}
u_{T}=-\partial_{x}P_{1}-P_{2}-\lambda u, \\[3pt]
v_{T}=-\alpha\sin^{2}\frac{v}{2}
+2h(u)\cos^{2}\frac{v}{2}
-2(P_{1}+\partial_{x}P_{2})\cos^{2}\frac{v}{2}-\lambda \sin v,\\[3pt]
\xi_{T}=\xi(\frac{\alpha}{2}+h(u)-P_{1}-\partial_{x}P_{2})\sin v
-2\lambda\xi\sin^{2}\frac{v}{2}
\end{array}
\right.
\end{array}
\eqno(3.1)
$$
with initial conditions given as
$$
\begin{array}{l}
\left\{\begin{array}{l}
u(0,Y)=u_{0}(x(0,Y)),\\[3pt]
v(0,Y)=2\arctan(u_{0,x}(x(0,Y))),\\[3pt]
\xi(0,x)=1,
\end{array}
\right.
\end{array}
\eqno(3.2)
$$
where $P_{1}, \partial_{x}P_{1}, P_{2}, \partial_{x}P_{2}$ are defined in (2.5).

We remark that the semi-linear system (3.1)-(3.2) is invariant under translation by
$2\pi$ in $v$. It would be more precise to use $e^{iv}$ as variable. For simplicity, we use
$v\in [-\pi, \pi]$ with endpoints identified.

Now we consider (3.1)-(3.2) as a system of ordinary differential equations on $(u, v, \xi)$ in the Banach
space
$$
X:=H^{1}(\mathbb{R})\times[L^{2}(\mathbb{R})\cap L^{\infty}(\mathbb{R})]\times L^{\infty}(\mathbb{R})
$$
with the norm
$$
\|(u,v,\xi)\|_{X}
=\|u\|_{H^{1}}
+\|v\|_{L^{2}}
+\|v\|_{L^{\infty}}
+\|\xi\|_{L^{\infty}}.
$$

From the standard ODE theory it follows that to obtain the local well-posedness of system (3.1)-(3.2), it suffices to prove that all functions on the right-hand side of (3.1) are locally Lipschitz continuous. \\

\noindent\textbf{Theorem 3.1.}  Given $u_{0}\in H^{1}(\mathbb{R})$,
there exist a $T_{0}>0$ such that the initial value problem (3.1)-(3.2) has a solution defined on $[0, T_{0}]$.\\

\noindent\textbf{Proof.} Our goal is to show that the right-hand side of (3.1) is Lipschitz
continuous in $(u,v,\xi)$ on every bounded domain $\Omega\subset X$ as follows
$$
\Omega
=\left\{(u,v,\xi):
\|u\|_{H^{1}}\leq A,~\|v\|_{L^{2}}\leq B,~
\|v\|_{L^{\infty}}\leq\frac{3\pi}{2},~
\xi(x)\in [C_{*}, C^{*}]~ a.e. x\in \mathbb{R}\right\}
$$
for some positive constants $A, B, C_{*}$ and $C^{*}$.

By the Sobolev inequality $\|u\|_{L^{\infty}}\leq \frac{1}{\sqrt{2}}\|u\|_{H^{1}}$
and the uniform bounds on $v, \xi$, it follows that the maps
$$
\begin{array}{l}
\lambda u,~-\alpha\sin^{2}\frac{v}{2}+2h(u)\cos^{2}\frac{v}{2},~
\lambda \sin v,~\xi(\frac{\alpha}{2}+h(u))\sin v,~
2\lambda\xi\sin^{2}\frac{v}{2}
\end{array}
$$
are all Lipschitz continuous from $\Omega$ into $L^{2}(\mathbb{R})\cap L^{\infty}(\mathbb{R})$. Our main task is to prove that the maps
$$
\begin{array}{l}
(u,v,\xi)\mapsto P_{i}, \quad (u,v,\xi)\mapsto \partial_{x}P_{i} \quad(i=1,2)
\end{array}
\eqno(3.3)
$$
are Lipschitz from $\Omega$ into $L^{2}(\mathbb{R})\cap L^{\infty}(\mathbb{R})$.
In fact, in what follows we can show that the above maps are Lipschitz from $\Omega$ into $H^{1}(\mathbb{R})$.

We first observe that for $(u,v,\xi)\in \Omega$ it holds
$$
\begin{array}{rl}
\textmd{meas}\{Y\in \mathbb{R};~|\frac{v(Y)}{2}|\geq \frac{\pi}{4}\}
&\leq \textmd{meas}\{Y\in \mathbb{R};~\sin^{2}\frac{v(Y)}{2}\geq \frac{1}{2}\}
\leq 2\int_{\{Y\in \mathbb{R};~\sin^{2}\frac{v(Y)}{2}\geq \frac{1}{2}\}}
\sin^{2}\frac{v(Y)}{2}dY\\[3pt]
&\leq \frac{1}{2}\int_{\{Y\in \mathbb{R};~\sin^{2}\frac{v(Y)}{2}\geq \frac{1}{2}\}}
v^{2}(Y)dY
\leq \frac{B^{2}}{2}.
\end{array}
$$
Thus, for any $\bar{Y}<Y$ we have
$$
\begin{array}{rl}
\int_{\bar{Y}}^{Y}\xi(s)\cos^{2}\frac{v(s)}{2}ds
\geq \int_{\{s\in [\bar{Y},Y];~|\frac{v(s)}{2}|\leq \frac{\pi}{4}\}}\frac{C_{*}}{2}ds
\geq \frac{C_{*}}{2}(Y-\bar{Y}-\frac{B^{2}}{2}).
\end{array}
$$
Introducing the exponentially decaying function
$$
\Theta_{1}(\zeta):=\min\left\{1, \exp\left(\frac{B^{2}C_{*}}{4}-\frac{C_{*}|\zeta|}{2}\right)\right\},
$$
we can check that
$$
\|\Theta_{1}\|_{L^{1}}
=(\int_{|\zeta|\leq \frac{B^{2}}{2}}+\int_{|\zeta|\geq\frac{B^{2}}{2}})
\Theta_{1}(\zeta)d\zeta
=B^{2}+\frac{4}{C_{*}}.
$$

Now we show that
$P_{i}, \partial_{x}P_{i}\in H^{1}(\mathbb{R})$ $(i=1,2)$, that is,
$$
P_{i},~\partial_{Y}P_{i},~\partial_{x}P_{i},~\partial_{Y}\partial_{x}P_{i}\in L^{2}(\mathbb{R}).
$$
Here we only consider the \textit{a priori} estimates for $\partial_{x}P_{i}$
and $\partial_{Y}\partial_{x}P_{i}$,
since the estimates for $P_{i}$ and $\partial_{Y}P_{i}$ are similar.
From the definition of $\partial_{x}P_{i}$, we have
$$
\begin{array}{rl}
&|\partial_{x}P_{1}(Y)|
\leq \frac{C^{*}}{2}
\left|\Theta_{1}*|h(u)\cos^{2}\frac{v}{2}+\frac{\alpha}{2}\sin^{2}\frac{v}{2}|(Y)\right|,\\[3pt]
&|\partial_{x}P_{2}(Y)|
\leq \frac{C^{*}}{2}
\left|\Theta_{1}*|ku\cos^{2}\frac{v}{2}|(Y)\right|.
\end{array}
$$
Since $h$ is locally Lipschitz continuous from $\mathbb{R}$ to $\mathbb{R}$ with $h(0)=0$, we have
$$
|h(u(T,Y))|=|h(u(T,Y))-h(0)|
\leq \sup_{|y|\leq A}|h^{\prime}(y)||u(T,Y)|
:=L|u(T,Y)|.
$$
By Young's inequality, we know
$$
\begin{array}{rl}
\|\partial_{x}P_{1}\|_{L^{2}}
&\leq \frac{C^{*}}{2}
\|\Theta_{1}\|_{L^{1}}
\|h(u)\cos^{2}\frac{v}{2}+\frac{\alpha}{2}\sin^{2}\frac{v}{2}\|_{L^{2}}
\leq \frac{C^{*}}{2}
\|\Theta_{1}\|_{L^{1}}(L\|u\|_{L^{2}}
+\frac{|\alpha|}{8}\|v\|_{L^{\infty}}\|v\|_{L^{2}})
<\infty,\\[3pt]
\|\partial_{x}P_{2}\|_{L^{2}}
&\leq \frac{C^{*}}{2}
\|\Theta_{1}\|_{L^{1}}\|ku\cos^{2}\frac{v}{2}\|_{L^{2}}
\leq\frac{C^{*}|k|}{2}
\|\Theta_{1}\|_{L^{1}}\|u\|_{L^{2}}<\infty.
\end{array}
\eqno(3.4)
$$
Moreover, differentiating $\partial_{x}P_{i}$ with respect to $Y$, we have
$$
\begin{array}{rl}
\partial_{Y}\partial_{x}P_{1}(Y)
&=-(\xi h(u)\cos^{2}\frac{v}{2}+\frac{\alpha}{2}\xi\sin^{2}\frac{v}{2})(Y)
+\frac{1}{2}(\int_{Y}^{+\infty}-\int_{-\infty}^{Y})
e^{-|\int_{\bar{Y}}^{Y}(\xi\cos^{2}\frac{v}{2})(T,\hat{Y})d\hat{Y}|}\cdot\\[3pt]
&\qquad
(\xi\cos^{2}\frac{v}{2})(Y)\textmd{sign}(\bar{Y}-Y)\cdot
(\xi h(u)\cos^{2}\frac{v}{2}+\frac{\alpha}{2}\xi\sin^{2}\frac{v}{2})(\bar{Y})d\bar{Y},\\[3pt]
\partial_{Y}\partial_{x}P_{2}(Y)
&=-k(\xi u\cos^{2}\frac{v}{2})(Y)
+\frac{k}{2}(\int_{Y}^{+\infty}-\int_{-\infty}^{Y})
e^{-|\int_{\bar{Y}}^{Y}(\xi\cos^{2}\frac{v}{2})(T,\hat{Y})d\hat{Y}|}\cdot\\[3pt]
&\qquad(\xi\cos^{2}\frac{v}{2})(Y)\textmd{sign}(\bar{Y}-Y)\cdot
(\xi u\cos^{2}\frac{v}{2})(\bar{Y})d\bar{Y}.
\end{array}
\eqno(3.5)
$$
Therefore,
$$
\begin{array}{rl}
|\partial_{Y}\partial_{x}P_{1}(Y)|
&\leq C^{*}(|h(u)|+\frac{|\alpha|}{8}v^{2})
+\frac{(C^{*})^{2}}{2}\left|\Theta_{1}*
|h(u)\cos^{2}\frac{v}{2}+\frac{\alpha}{2}\xi\sin^{2}\frac{v}{2}|\right|,\\[3pt]
|\partial_{Y}\partial_{x}P_{2}(Y)|
&\leq|k|C^{*}|u|
+\frac{|k|(C^{*})^{2}}{2}\left|\Theta_{1}*|u\cos^{2}\frac{v}{2}|\right|.
\end{array}
$$
Similar to (3.4), we can get $\|\partial_{Y}\partial_{x}P_{i}\|_{L^{2}}<\infty$ $(i=1,2)$.

Next, we establish the Lipschitz continuity of the map given in (3.3). It is suffices to show that the partial derivatives
$$
\partial_{u}P_{i},~\partial_{v}P_{i},~\partial_{\xi}P_{i},~
\partial_{u}\partial_{x}P_{i},~\partial_{v}\partial_{x}P_{i},~
\partial_{\xi}\partial_{x}P_{i}
$$
are uniformly bounded linear operators from the appropriate spaces into $H^{1}(\mathbb{R})$. Here we only consider $\partial_{u}\partial_{x}P_{i}$, since the other partial derivatives
can be handled similarly.

For a given $(u,v,\xi)\in \Omega$ and a test function $\phi\in H^{1}(\mathbb{R})$,
the operators $\partial_{u}(\partial_{x}P_{i})$ and $\partial_{u}(\partial_{Y}\partial_{x}P_{i})$ are defined as follows.
$$
\begin{array}{rl}
[\partial_{u}(\partial_{x}P_{1})\cdot\phi](Y)
&=\frac{1}{2}(\int_{Y}^{+\infty}-\int_{-\infty}^{Y})
e^{-|\int_{\bar{Y}}^{Y}(\xi\cos^{2}\frac{v}{2})(T,\hat{Y})d\hat{Y}|}
(\phi\xi h^{\prime}(u)\cos^{2}\frac{v}{2})(\bar{Y})d\bar{Y},\\[5pt]
[\partial_{u}(\partial_{x}P_{2})\cdot\phi](Y)
&=\frac{k}{2}(\int_{Y}^{+\infty}-\int_{-\infty}^{Y})
e^{-|\int_{\bar{Y}}^{Y}(\xi\cos^{2}\frac{v}{2})(T,\hat{Y})d\hat{Y}|}
(\phi\xi\cos^{2}\frac{v}{2})(\bar{Y})d\bar{Y},\\[5pt]
[\partial_{u}(\partial_{Y}\partial_{x}P_{1})\cdot\phi](Y)
&=-(\phi\xi h^{\prime}(u)\cos^{2}\frac{v}{2})(Y)
+\frac{1}{2}(\int_{Y}^{+\infty}-\int_{-\infty}^{Y})
e^{-|\int_{\bar{Y}}^{Y}(\xi\cos^{2}\frac{v}{2})(T,\hat{Y})d\hat{Y}|}\cdot\\[5pt]
&\quad
(\xi\cos^{2}\frac{v}{2})(Y)\textmd{sign}(\bar{Y}-Y)\cdot
(\phi\xi h^{\prime}(u)\cos^{2}\frac{v}{2})(\bar{Y})d\bar{Y},\\[5pt]
[\partial_{u}(\partial_{Y}\partial_{x}P_{2})\cdot\phi](Y)
&=-k(\phi\xi \cos^{2}\frac{v}{2})(Y)
+\frac{k}{2}(\int_{Y}^{+\infty}-\int_{-\infty}^{Y})
e^{-|\int_{\bar{Y}}^{Y}(\xi\cos^{2}\frac{v}{2})(T,\hat{Y})d\hat{Y}|}\cdot\\[5pt]
&\quad
(\xi\cos^{2}\frac{v}{2})(Y)\textmd{sign}(\bar{Y}-Y)\cdot
(\phi\xi\cos^{2}\frac{v}{2})(\bar{Y})d\bar{Y}.
\end{array}
$$
Thus,
$$
\begin{array}{rl}
\|\partial_{u}(\partial_{x}P_{1})\cdot\phi\|_{L^{2}}
&\leq \frac{C^{*}}{2}\|h^{\prime}(u)\|_{L^{\infty}}
\big\|\Theta_{1}*|\phi|\big\|_{L^{2}}
\leq \frac{C^{*}}{2}\|h^{\prime}(u)\|_{L^{\infty}}
\|\Theta_{1}\|_{L^{1}}\|\phi\|_{L^{2}}\\[3pt]
&\leq \frac{C^{*}}{2}\|h^{\prime}(u)\|_{L^{\infty}}
\|\Theta_{1}\|_{L^{1}}\|\phi\|_{H^{1}},\\[3pt]
\|\partial_{u}(\partial_{x}P_{2})\cdot\phi\|_{L^{2}}
&\leq \frac{|k|C^{*}}{2}\big\|\Theta_{1}*|\phi|\big\|_{L^{2}}
\leq \frac{|k|C^{*}}{2}\|\Theta_{1}\|_{L^{1}}\|\phi\|_{H^{1}},\\[3pt]
\|\partial_{u}(\partial_{Y}\partial_{x}P_{1})\cdot\phi\|_{L^{2}}
&\leq C^{*}\|h^{\prime}(u)\|_{L^{\infty}}\|\phi\|_{L^{2}}
+\frac{(C^{*})^{2}}{2}\|h^{\prime}(u)\|_{L^{\infty}}
\big\|\Theta_{1}*|\phi|\big\|_{L^{2}}\\[3pt]
&\leq C^{*}\|h^{\prime}(u)\|_{L^{\infty}}\|\phi\|_{H^{1}}
+\frac{(C^{*})^{2}}{2}\|h^{\prime}(u)\|_{L^{\infty}}
\|\Theta_{1}\|_{L^{1}}\|\phi\|_{H^{1}},\\[3pt]
\|\partial_{u}(\partial_{Y}\partial_{x}P_{2})\cdot\phi\|_{L^{2}}
&\leq |k|C^{*}\|\phi\|_{L^{2}}
+\frac{|k|(C^{*})^{2}}{2}
\left\|\Theta_{1}*|\phi|\right\|_{L^{2}}
\leq |k|C^{*}\|\phi\|_{H^{1}}
+\frac{|k|(C^{*})^{2}}{2}
\|\Theta_{1}\|_{L^{1}}\|\phi\|_{H^{1}}.
\end{array}
$$
Hence we obtain that $\partial_{u}\partial_{x}P_{1}$ is a bounded linear operator from
$H^{1}(\mathbb{R})$ to $H^{1}(\mathbb{R})$. As above, we can bound the other partial
derivatives, thus the uniform Lipschitz continuous of the map in (3.3) is now verified.
Then using the standard ODE theory in the Banach space, the local existence of a solution
to the Cauchy problem (3.1)-(3.2) is established, that is, the initial problem admits
a unique solution on $[0,T_{0}]$ for some $T_{0}>0$. \hfill $\Box$\\

Next, we shall prove that the local solution for (3.1)-(3.2) can be extended to the global one. \\

\noindent\textbf{Theorem 3.2.}  If $u_{0}\in H^{1}(\mathbb{R})$,
then the Cauchy problem (3.1)-(3.2) has a unique solution defined for all $T\geq0$.\\

\noindent\textbf{Proof.} In view of the proof of Theorem 3.1, to extend the local solution we only need to show that the quantity
$$
\|u\|_{H^{1}}+\|v\|_{L^{2}}+\|v\|_{L^{\infty}}
+\|\xi\|_{L^{\infty}}+\|\frac{1}{\xi}\|_{L^{\infty}}
$$
is uniformly bounded on any bounded time interval.

As long as the local solution of (3.1)-(3.2) is defined, we claim that
$$
\begin{array}{rl}
u_{Y}=u_{x}x_{Y}
=u_{x}\frac{\xi}{1+u_{x}^{2}}
=\frac{1}{2}\xi\sin v.
\end{array}
\eqno(3.6)
$$
In fact, from (3.1) we have
$$
\begin{array}{rl}
(u_{Y})_{T}=(u_{T})_{Y}
&=-\partial_{Y}\partial_{x}P_{1}
-\partial_{Y}P_{2}-\lambda u_{Y}
=\frac{-\partial_{x}^{2}P_{1}
-\partial_{x}P_{2}-\lambda u_{x}}{Y_{x}}\\[3pt]
&=\xi\left[h(u)\cos^{2}\frac{v}{2}
+\frac{\alpha}{2}\sin^{2}\frac{v}{2}
-(P_{1}+\partial_{x}P_{2})\cos^{2}\frac{v}{2}
-\frac{1}{2}\lambda\sin v\right].
\end{array}
\eqno(3.7)
$$
Moreover, from (3.1) and (3.6) we have
$$
\begin{array}{rl}
(\frac{1}{2}\xi\sin v)_{T}
&=\frac{1}{2}\xi_{T}\sin v
+\frac{1}{2}\xi v_{T}\cos v
=\frac{1}{2}\sin v[\xi(\frac{\alpha}{2}+h(u)-P_{1}-\partial_{x}P_{2})\sin v
-2\lambda\xi\sin^{2}\frac{v}{2}]\\[3pt]
&\quad+\frac{1}{2}\xi\cos v[-\alpha\sin^{2}\frac{v}{2}
+2h(u)\cos^{2}\frac{v}{2}
-2\cos^{2}\frac{v}{2}(P_{1}+\partial_{x}P_{2})-\lambda \sin v]\\[3pt]
&=\xi\left[h(u)\cos^{2}\frac{v}{2}
+\frac{\alpha}{2}\sin^{2}\frac{v}{2}
-(P_{1}+\partial_{x}P_{2})\cos^{2}\frac{v}{2}
-\frac{1}{2}\lambda\sin v\right]
=(u_{Y})_{T}.
\end{array}
$$
When $T=0$, we have
$$
u_{Y}(0,Y)=\frac{1}{2}\sin v(0,Y), \quad
\xi(0,Y)=1,
$$
then the claim holds for all $T\geq 0$, as long as the solution is defined.

We denote the energy in the new coordinate by
$E(T):=\int_{\mathbb{R}}(u^{2}\xi\cos^{2}\frac{v}{2}+\xi\sin^{2}\frac{v}{2})dY$.
From (3.1), we have
$$
\begin{array}{rl}
&\frac{d}{dT}\int_{\mathbb{R}}(u^{2}\xi\cos^{2}\frac{v}{2}+\xi\sin^{2}\frac{v}{2})dY\\[3pt]
&=\int_{\mathbb{R}}[(u^{2}\cos^{2}\frac{v}{2}+\sin^{2}\frac{v}{2})\xi_{T}
+2u\xi u_{T}\cos^{2}\frac{v}{2}+\xi(1-u^{2})v_{T}\sin\frac{v}{2}\cos\frac{v}{2}]dY\\[3pt]
&=\int_{\mathbb{R}}[\alpha\xi u^{2}\sin\frac{v}{2}\cos\frac{v}{2}
+2\xi h(u)\sin\frac{v}{2}\cos\frac{v}{2}
-2\xi(P_{1}+\partial_{x}P_{2})\sin\frac{v}{2}\cos\frac{v}{2}\\[3pt]
&\quad-2u\xi(\partial_{x}P_{1}+P_{2})\cos^{2}\frac{v}{2}
-2\lambda\xi\sin^{2}\frac{v}{2}-2\lambda\xi u^{2}\cos^{2}\frac{v}{2}]dY\\[3pt]
&=\alpha\int_{\mathbb{R}}u^{2}u_{Y}dY
+2\int_{\mathbb{R}}h(u)u_{Y}dY
-2\int_{\mathbb{R}}[u(P_{1}+\partial_{x}P_{2})]_{Y}dY\\[3pt]
&\quad-2k\int_{\mathbb{R}} u^{2}\xi\cos^{2}\frac{v}{2}dY
-2\lambda\int_{\mathbb{R}}(u^{2}\xi\cos^{2}\frac{v}{2}+\xi\sin^{2}\frac{v}{2})dY\\[3pt]
&=-2k\int_{\mathbb{R}} u^{2}\xi\cos^{2}\frac{v}{2}dY
-2\lambda\int_{\mathbb{R}}(u^{2}\xi\cos^{2}\frac{v}{2}+\xi\sin^{2}\frac{v}{2})dY\\[3pt]
&\leq 2(|k|+|\lambda|)\int_{\mathbb{R}}(u^{2}\xi\cos^{2}\frac{v}{2}+\xi\sin^{2}\frac{v}{2})dY.
\end{array}
$$
By Gronwall's inequality, we can deduce that
$$
E(T)\leq e^{2(|k|+|\lambda|)T}E(0).
$$

Assume $T\in [0, \tilde{T}]$ with any fixed $\tilde{T}>0$. As long as the solution exists at some $T\in [0, \tilde{T}]$, we can obtain
$$
\begin{array}{rl}
E(T)\leq e^{2(|k|+|\lambda|)\tilde{T}}E(0):=\mathfrak{C},
\end{array}
\eqno(3.8)
$$
and
$$
\begin{array}{rl}
\sup_{Y\in \mathbb{R}}|u^{2}(T,Y)|
\leq 2\int_{\mathbb{R}}|uu_{Y}|dY
=2\int_{\mathbb{R}}|u\sin\frac{v}{2}\cos\frac{v}{2}|\xi dY
\leq \int_{\mathbb{R}}(u^{2}\cos^{2}\frac{v}{2}+\sin^{2}\frac{v}{2})\xi dY
\leq \mathfrak{C},
\end{array}
$$
which implies
$$
\begin{array}{rl}
\|u(T)\|_{L^{\infty}}\leq \mathfrak{C}^{1/2}.
\end{array}
\eqno(3.9)
$$

Since $h$ is locally Lipschitz continuous from $\mathbb{R}$ to $\mathbb{R}$ with $h(0)=0$, as long as the solution exists at some $T\in [0, \tilde{T}]$, we have
$$
|h(u(T,Y))|=|h(u(T,Y))-h(0)|
\leq \sup_{|y|\leq \mathfrak{C}^{1/2}}|h^{\prime}(y)||u(T,Y)|
:=L|u(T,Y)|.
$$
From (2.5), we have
$$
\begin{array}{rl}
&|P_{1}(T,Y)|\\[3pt]
&=|\frac{1}{2}\int_{-\infty}^{+\infty}
e^{-|\int_{\bar{Y}}^{Y}(\xi\cos^{2}\frac{v}{2})(T,\hat{Y})d\hat{Y}|}\cdot
(\xi h(u)\cos^{2}\frac{v}{2}
+\frac{\alpha}{2}\xi\sin^{2}\frac{v}{2})(T,\bar{Y})d\bar{Y}|\\[3pt]
&\leq \frac{L}{2}\int_{-\infty}^{+\infty}
e^{-|\int_{\bar{Y}}^{Y}(\xi\cos^{2}\frac{v}{2})(T,\hat{Y})d\hat{Y}|}\cdot
(\xi |u|\cos^{2}\frac{v}{2})(\bar{Y})d\bar{Y}
+\frac{|\alpha|}{4}
\int_{-\infty}^{+\infty}\xi\sin^{2}\frac{v}{2}d\bar{Y}\\[3pt]
&\leq \frac{L}{4}\int_{-\infty}^{+\infty}
e^{-|\int_{\bar{Y}}^{Y}(\xi\cos^{2}\frac{v}{2})(T,\hat{Y})d\hat{Y}|}\cdot
\xi (\cos^{2}\frac{v}{2}+u^{2}\cos^{2}\frac{v}{2})d\bar{Y}
+\frac{|\alpha|}{4}
\int_{-\infty}^{+\infty}\xi\sin^{2}\frac{v}{2}d\bar{Y}\\[3pt]
&\leq \frac{L}{4}\int_{-\infty}^{+\infty}
e^{-|\int_{\bar{Y}}^{Y}(\xi\cos^{2}\frac{v}{2})(T,\hat{Y})d\hat{Y}|}\cdot
\xi \cos^{2}\frac{v}{2}d\bar{Y}
+\frac{L}{4}\int_{-\infty}^{+\infty}\xi u^{2}\cos^{2}\frac{v}{2}d\bar{Y}
+\frac{|\alpha|}{4}
\int_{-\infty}^{+\infty}\xi\sin^{2}\frac{v}{2}d\bar{Y}.
\end{array}
$$
For the first term in the right hand,
$$
\begin{array}{rl}
&\int_{-\infty}^{+\infty}
e^{-|\int_{\bar{Y}}^{Y}(\xi\cos^{2}\frac{v}{2})(T,\hat{Y})d\hat{Y}|}\cdot
\xi \cos^{2}\frac{v}{2}d\bar{Y}\\[3pt]
&=(\int_{-\infty}^{Y}
+\int_{Y}^{+\infty})
e^{-|\int_{\bar{Y}}^{Y}(\xi\cos^{2}\frac{v}{2})(T,\hat{Y})d\hat{Y}|}\cdot
\xi \cos^{2}\frac{v}{2}d\bar{Y}\\[3pt]
&=\int_{-\infty}^{Y}
e^{-\int_{\bar{Y}}^{Y}(\xi\cos^{2}\frac{v}{2})(T,\hat{Y})d\hat{Y}}\cdot
\xi \cos^{2}\frac{v}{2}d\bar{Y}
+\int_{Y}^{+\infty}
e^{-\int_{Y}^{\bar{Y}}(\xi\cos^{2}\frac{v}{2})(T,\hat{Y})d\hat{Y}}\cdot
\xi \cos^{2}\frac{v}{2}d\bar{Y}\\[3pt]
&=\int_{-\infty}^{Y}\frac{d}{d\bar{Y}}
(e^{-\int_{\bar{Y}}^{Y}(\xi\cos^{2}\frac{v}{2})(T,\hat{Y})d\hat{Y}})d\bar{Y}
+\int_{Y}^{+\infty}-\frac{d}{d\bar{Y}}
(e^{-\int_{Y}^{\bar{Y}}(\xi\cos^{2}\frac{v}{2})(T,\hat{Y})d\hat{Y}})d\bar{Y}\\[3pt]
&=2-e^{-\int_{-\infty}^{Y}(\xi\cos^{2}\frac{v}{2})(T,\hat{Y})d\hat{Y}}
-e^{-\int_{Y}^{+\infty}(\xi\cos^{2}\frac{v}{2})(T,\hat{Y})d\hat{Y}}\\[3pt]
&\leq 2.
\end{array}
$$
Thus,
$$
\begin{array}{rl}
\|P_{1}(T)\|_{L^{\infty}}
\leq \frac{L}{2}+\frac{L+|\alpha|}{4}\mathfrak{C}.
\end{array}
\eqno(3.10)
$$
For $\partial_{x}P_{2}$,
$$
\begin{array}{rl}
|\partial_{x}P_{2}(T,Y)|
&=|\frac{k}{2}(\int_{Y}^{+\infty}-\int_{-\infty}^{Y})
e^{-|\int_{\bar{Y}}^{Y}(\xi\cos^{2}\frac{v}{2})(T,\hat{Y})d\hat{Y}|}
(\xi u\cos^{2}\frac{v}{2})(T,\bar{Y})d\bar{Y}|\\[3pt]
&\leq \frac{|k|}{2}(\int_{Y}^{+\infty}+\int_{-\infty}^{Y})
e^{-|\int_{\bar{Y}}^{Y}(\xi\cos^{2}\frac{v}{2})(T,\hat{Y})d\hat{Y}|}
(\xi |u|\cos^{2}\frac{v}{2})(T,\bar{Y})d\bar{Y}|\\[3pt]
&\leq \frac{|k|}{4}
(\int_{Y}^{+\infty}+\int_{-\infty}^{Y})
e^{-|\int_{\bar{Y}}^{Y}(\xi\cos^{2}\frac{v}{2})(T,\hat{Y})d\hat{Y}|}
\xi(\cos^{2}\frac{v}{2}+u^{2}\cos^{2}\frac{v}{2})(T,\bar{Y})d\bar{Y}\\[3pt]
&=\frac{|k|}{4}
\int_{-\infty}^{+\infty}
e^{-|\int_{\bar{Y}}^{Y}(\xi\cos^{2}\frac{v}{2})(T,\hat{Y})d\hat{Y}|}
\xi(\cos^{2}\frac{v}{2}+u^{2}\cos^{2}\frac{v}{2})(T,\bar{Y})d\bar{Y}\\[3pt]
&\leq \frac{|k|}{2}+\frac{|k|}{4}\mathfrak{C},
\end{array}
$$
that is,
$$
\begin{array}{rl}
\|\partial_{x}P_{2}(T)\|_{L^{\infty}}
\leq \frac{|k|}{2}+\frac{|k|}{4}\mathfrak{C}.
\end{array}
\eqno(3.11)
$$

With the estimates (3.9)-(3.11), it is now clear from the third equation in (3.1) that
$$
\begin{array}{rl}
|\xi_{T}|
&\leq (\frac{|\alpha|}{2}+L\|u(T)\|_{L^{\infty}}
+\|P_{1}(T)\|_{L^{\infty}}+\|\partial_{x}P_{2}(T)\|_{L^{\infty}})\xi\\[3pt]
&\leq (\frac{|\alpha|}{2}+L\mathfrak{C}^{1/2}
+\frac{L}{2}+\frac{L+|\alpha|}{4}\mathfrak{C}
+\frac{|k|}{2}+\frac{|k|}{4}\mathfrak{C})\xi\\[3pt]
&:=\mathfrak{D}_{1}\xi,
\end{array}
$$
as long as the solution exists at some $T\in [0, \tilde{T}]$.

Since $\xi(0,Y)=1$, we know
$$
\begin{array}{rl}
e^{-\mathfrak{D}_{1}\tilde{T}}\leq e^{-\mathfrak{D}_{1}T}
\leq \xi(T)\leq e^{\mathfrak{D}_{1}T}\leq e^{\mathfrak{D}_{1}\tilde{T}}.
\end{array}
\eqno(3.12)
$$
Similarly, we can get the estimate for $v_{T}$ by the second equation in (3.1)
$$
\begin{array}{rl}
|v_{T}|
&\leq |\alpha|+2L\|u(T)\|_{L^{\infty}}+2\|P_{1}(T)\|_{L^{\infty}}
+2\|\partial_{x}P_{2}(T)\|_{L^{\infty}}+|\lambda|\\[3pt]
&\leq |\alpha|+2L\mathfrak{C}^{1/2}
+L+\frac{L+|\alpha|}{2}\mathfrak{C}
+|k|+\frac{|k|}{2}\mathfrak{C}+|\lambda|\\[3pt]
&:=\mathfrak{D}_{2}.
\end{array}
$$
Hence,
$$
\|v(T)\|_{L^{\infty}}
\leq \|v(0)\|_{L^{\infty}}+\mathfrak{D}_{2}T
\leq \|v(0)\|_{L^{\infty}}+\mathfrak{D}_{2}\tilde{T}.
$$

Moreover, the first equation in (3.1) implies
$$
\frac{d}{dT}\int_{\mathbb{R}}u^{2}dY
=2\int_{\mathbb{R}}uu_{T}dY
=2\int_{\mathbb{R}}u(-\partial_{x}P_{1}-P_{2})dY-2\lambda\int_{\mathbb{R}}u^{2}dY
$$
and
$$
\frac{d}{dT}\int_{\mathbb{R}}(\partial_{Y}u)^{2}dY
=2\int_{\mathbb{R}}\partial_{Y}u\partial_{Y}u_{T}dY
=2\int_{\mathbb{R}}\partial_{Y}u(-\partial_{Y}\partial_{x}P_{1}-\partial_{Y}P_{2})dY
-2\lambda\int_{\mathbb{R}}(\partial_{Y}u)^{2}dY.
$$
Thus,
$$
\frac{d}{dT}(e^{2\lambda T}\|u(T)\|_{L^{2}}^{2})
\leq 2e^{2\lambda T}\|u(T)\|_{L^{2}}
(\|\partial_{x}P_{1}(T)\|_{L^{2}}+\|P_{2}(T)\|_{L^{2}})
$$
and
$$
\frac{d}{dT}(e^{2\lambda T}\|\partial_{Y}u(T)\|_{L^{2}}^{2})
\leq 2e^{2\lambda T}\|\partial_{Y}u(T)\|_{L^{2}}
(\|\partial_{Y}\partial_{x}P_{1}(T)\|_{L^{2}}+\|\partial_{Y}P_{2}(T)\|_{L^{2}}),
$$
which implies that
$$
\begin{array}{rl}
&\|u(T)\|_{H^{1}}\\[3pt]
&\leq e^{-\lambda T}\|u_{0}\|_{H^{1}}
+e^{-\lambda T}\int_{0}^{T}e^{\lambda s}
(\|\partial_{x}P_{1}(s)\|_{L^{2}}+\|P_{2}(s)\|_{L^{2}}
+\|\partial_{Y}\partial_{x}P_{1}(s)\|_{L^{2}}+\|\partial_{Y}P_{2}(s)\|_{L^{2}})ds\\[3pt]
&\leq \|u_{0}\|_{H^{1}}
+\int_{0}^{\tilde{T}}e^{\lambda s}
(\|\partial_{x}P_{1}(s)\|_{L^{2}}+\|P_{2}(s)\|_{L^{2}}
+\|\partial_{Y}\partial_{x}P_{1}(s)\|_{L^{2}}+\|\partial_{Y}P_{2}(s)\|_{L^{2}})ds.
\end{array}
$$
Next, we estimate $\|\partial_{x}P_{1}(s)\|_{L^{2}},\|P_{2}(s)\|_{L^{2}},
\|\partial_{Y}\partial_{x}P_{1}(s)\|_{L^{2}}$ and $\|\partial_{Y}P_{2}(s)\|_{L^{2}}$.
For the estimate on $\|\partial_{Y}\partial_{x}P_{1}(s)\|_{L^{2}}$,
we first look for a lower bound of $|\int_{\bar{Y}}^{Y}(\xi\cos^{2}\frac{v}{2})(s,\hat{Y})d\hat{Y}|$.
We denote by $\Lambda$ the right-hand side of (3.12), so that
$\Lambda^{-1}\leq\xi(T)\leq \Lambda$. For $Y>\bar{Y}$,
$$
\begin{array}{rl}
\int_{\bar{Y}}^{Y}(\xi\cos^{2}\frac{v}{2})(s,\hat{Y})d\hat{Y}
&\geq \int_{\{\hat{Y}\in[\bar{Y},Y], |\frac{v(\hat{Y})}{2}|\leq\frac{\pi}{4}\}}(\xi\cos^{2}\frac{v}{2})(s,\hat{Y})d\hat{Y}
\geq \frac{1}{2}\int_{\{\hat{Y}\in[\bar{Y},Y], |\frac{v(\hat{Y})}{2}|\leq\frac{\pi}{4}\}}\xi(s,\hat{Y})d\hat{Y}\\[3pt]
&\geq \frac{\Lambda^{-1}}{2}(Y-\bar{Y})
-\frac{1}{2}\int_{\{\hat{Y}\in[\bar{Y},Y], |\frac{v(\hat{Y})}{2}|\geq\frac{\pi}{4}\}}\xi(s,\hat{Y})d\hat{Y}\\[3pt]
&\geq \frac{\Lambda^{-1}}{2}(Y-\bar{Y})
-\int_{\{\hat{Y}\in[\bar{Y},Y], |\frac{v(\hat{Y})}{2}|\geq\frac{\pi}{4}\}}(\xi\sin^{2}\frac{v}{2})(s,\hat{Y})d\hat{Y}\\[3pt]
&\geq \frac{\Lambda^{-1}}{2}(Y-\bar{Y})-\mathfrak{C}.
\end{array}
$$
We define
$$
\Theta_{2}(\zeta):=
\min\left\{1, \exp\left(\mathfrak{C}-\frac{\Lambda^{-1}|\zeta|}{2}\right)\right\}
$$
with the property that
$$
\|\Theta_{2}\|_{L^{1}}
=4\Lambda(\mathfrak{C}+1)
=4(\mathfrak{C}+1)e^{\mathfrak{D}_{1}\tilde{T}}.
$$
Hence, from (3.5), we have
$$
\begin{array}{rl}
\|\partial_{Y}\partial_{x}P_{1}(s)\|_{L^{2}}
&\leq \|\xi h(u)\cos^{2}\frac{v}{2}+\frac{\alpha}{2}\xi\sin^{2}\frac{v}{2}\|_{L^{2}}
+\frac{1}{2}\|\xi\|_{L^{\infty}}\left\|\Theta_{2}*|\xi h(u)\cos^{2}\frac{v}{2}+\frac{\alpha}{2}\xi\sin^{2}\frac{v}{2}|\right\|_{L^{2}}\\[3pt]
&\leq (1+\frac{1}{2}\|\xi\|_{L^{\infty}}\|\Theta_{2}\|_{L^{1}})
\|\xi h(u)\cos^{2}\frac{v}{2}+\frac{\alpha}{2}\xi\sin^{2}\frac{v}{2}\|_{L^{2}}\\[3pt]
&\leq (1+\frac{1}{2}\|\xi\|_{L^{\infty}}\|\Theta_{2}\|_{L^{1}})
(L\|\xi u\cos^{2}\frac{v}{2}\|_{L^{2}}+\|\frac{\alpha}{2}\xi\sin^{2}\frac{v}{2}\|_{L^{2}})\\[3pt]
&\leq (1+\frac{1}{2}\|\xi\|_{L^{\infty}}\|\Theta_{2}\|_{L^{1}})
(L\|\xi\|_{L^{\infty}}^{\frac{1}{2}}\|\xi^{\frac{1}{2}} u\cos\frac{v}{2}\|_{L^{2}}
+\frac{|\alpha|}{2}\|\xi\|_{L^{\infty}}^{\frac{1}{2}}\|\xi^{\frac{1}{2}}\sin\frac{v}{2}\|_{L^{2}})\\[3pt]
&\leq (L+\frac{|\alpha|}{2})(1+\frac{1}{2}\|\xi\|_{L^{\infty}}\|\Theta_{2}\|_{L^{1}})
\|\xi\|_{L^{\infty}}^{\frac{1}{2}}E(T)^{\frac{1}{2}}\\[3pt]
&\leq (L+\frac{|\alpha|}{2})
\mathfrak{C}^{\frac{1}{2}}e^{\frac{1}{2}\mathfrak{D}_{1}\tilde{T}}
[1+2(\mathfrak{C}+1)e^{2\mathfrak{D}_{1}\tilde{T}}].
\end{array}
$$
The estimates for $\|\partial_{x}P_{1}(s)\|_{L^{2}},\|P_{2}(s)\|_{L^{2}}$ and $\|\partial_{Y}P_{2}(s)\|_{L^{2}}$ are entirely similar. This establishes the uniformly
boundedness of $\|u(T)\|_{H^{1}}$ as long as the solution exists at some $T\in [0, \tilde{T}]$.

Lastly, we try to bound $\|v\|_{L^{2}}$. Note that $v\in [-\pi,\pi]$ and the inequality $|\sin x|\leq |x|$ for $x\in \mathbb{\mathbb{R}}$. Multiplying $v$ to the second equation in (3.1) and integrating with respect to $Y\in \mathbb{R}$, we have
$$
\begin{array}{rl}
&\frac{1}{2}\frac{d}{dT}\|v(T)\|_{L^{2}}^{2}\\[3pt]
&=-\alpha\int_{\mathbb{R}}(v\sin^{2}\frac{v}{2})(T,Y)d Y
+2\int_{\mathbb{R}}(vh(u)\cos^{2}\frac{v}{2})(T,Y)d Y\\[3pt]
&\quad-2\int_{\mathbb{R}}(v(P_{1}+\partial_{x}P_{2})\cos^{2}\frac{v}{2})(T,Y)d Y
-\lambda \int_{\mathbb{R}}(v\sin v)(T,Y)d Y\\[3pt]
&\leq \frac{|\alpha|}{2}\|v(T)\|_{L^{2}}^{2}
+2L\|v(T)\|_{L^{2}}\|u(T)\|_{L^{2}}
+2\|v(T)\|_{L^{2}}\|(P_{1}+\partial_{x}P_{2})(T)\|_{L^{2}}
+|\lambda| \|v(T)\|_{L^{2}}^{2},
\end{array}
$$
that is,
$$
\frac{d}{dT}\|v(T)\|_{L^{2}}
\leq (\frac{|\alpha|}{2}+|\lambda|)\|v(T)\|_{L^{2}}
+2(L\|u(T)\|_{L^{2}}+\|(P_{1}+\partial_{x}P_{2})(T)\|_{L^{2}}).
$$
By the previous bounds, it is clear that $\|v(T)\|_{L^{2}}$ is uniformly bounded as long as the solution exists at some $T\in [0, \tilde{T}]$. This completes the proof of Theorem 3.2.  \hfill $\Box$\\

For future use, we give an important property of the above global solution.\\

\noindent\textbf{Lemma 3.3.} Consider the set of times
$$
\mathcal{N}:=
\left\{T\geq 0;~
\textmd{meas}\{Y\in\mathbb{R};~ v(T, Y)=-\pi\}>0\right\}.
$$
Then
$$
\textmd{meas}(\mathcal{N})=0.
$$

\noindent\textbf{Proof.}
Since the Lebesgue measure is $\sigma$-finite, it is suffices to prove that
$\textmd{meas}(\mathcal{N}\cap I)=0$ for any compact interval $I$ of $\mathbb{R}_{+}$.

Similar to the proof of Theorem 3.2, we can show that there exists a constant
$M$ depending on $I$ such that
$$
\|(h(u)-P_{1}-\partial_{x}P_{2})(T)\|_{L^{\infty}}\leq M, \quad\forall~T\in I.
$$
Taking $0<\delta\ll 1$ such that
$$
M\delta+\sqrt{2}\lambda\delta^{\frac{1}{2}}
<\frac{|\alpha|}{2}(1-\frac{\delta}{2}).
$$
By the second equation in (3.1), we have
$$
\begin{array}{rl}
&|\alpha|\sin^{2}\frac{v}{2}
-\left(\|(h(u)-P_{1}-\partial_{x}P_{2})(T)\|_{L^{\infty}}\cos^{2}\frac{v}{2}
+2|\lambda| \cos \frac{v}{2}\right)\\[3pt]
&\leq|v_{T}|\leq |\alpha|\sin^{2}\frac{v}{2}+
\left(\|(h(u)-P_{1}-\partial_{x}P_{2})(T)\|_{L^{\infty}}\cos^{2}\frac{v}{2}
+2|\lambda| \cos \frac{v}{2}\right).
\end{array}
$$
Thus, whenever $1+\cos v(T,Y)<\delta$ (which means $1-\frac{\delta}{2}<\sin^{2}\frac{v}{2}\leq1$ and $\cos \frac{v}{2}\leq \sqrt{\frac{\delta}{2}}$),
we have
$$
\begin{array}{rl}
\frac{|\alpha|}{4}
<\frac{|\alpha|}{2}(1-\frac{\delta}{2})
<|v_{T}|<\frac{3|\alpha|}{2}(1-\frac{\delta}{6})
<\frac{3|\alpha|}{2}.
\end{array}
\eqno(3.13)
$$

For any fixed $T\in I$, we define
$$
\mathcal{O}(T,\delta):=\{Y\in \mathbb{R};~ 1+\cos v(T,Y)<\delta\}.
$$
We claim that $\textmd{meas} (\mathcal{O}(T,\delta))$ is finite. Indeed, by the second equation in (3.1), we have
$$
\begin{array}{rl}
&\int_{\mathcal{O}(T,\delta)}v_{T}^{2}d Y\\[3pt]
&=\int_{\mathcal{O}(T,\delta)}[\alpha^{2}\sin^{4}\frac{v}{2}+4(h(u)
-P_{1}-\partial_{x}P_{2})^{2}\cos^{4}\frac{v}{2}
+\lambda^{2}\sin^{2}v\\[3pt]
&\quad-(h(u)-P_{1}-\partial_{x}P_{2})(4\alpha\sin^{2}\frac{v}{2}\cos^{2}\frac{v}{2}
+2\lambda\sin v\cos^{2}\frac{v}{2})
+2\lambda\alpha\sin v\sin^{2}\frac{v}{2}]dY\\[3pt]
&\leq \int_{\mathcal{O}(T,\delta)}[\frac{\alpha^{2}}{4}v^{2}
+4(h(u)-P_{1}-\partial_{x}P_{2})^{2}
+\lambda^{2}v^{2}+2(|\alpha|+\lambda)|h(u)-P_{1}-\partial_{x}P_{2}||v|
+\frac{1}{2}|\lambda\alpha|v^{2}]dY\\[3pt]
&\leq (\frac{\alpha^{2}}{4}+\lambda^{2}+\frac{1}{2}|\lambda\alpha|)
\int_{\mathcal{O}(T,\delta)}v^{2}(T,Y)d Y
+4\int_{\mathcal{O}(T,\delta)}(h(u)-P_{1}-\partial_{x}P_{2})^{2}d Y\\[3pt]
&\quad+2(|\alpha|+|\lambda|)\int_{\mathcal{O}(T,\delta)}
|h(u)-P_{1}-\partial_{x}P_{2}||v|d Y\\[3pt]
&\leq (\frac{\alpha^{2}}{4}+\lambda^{2}+\frac{1}{2}|\lambda\alpha|)\|v(T)\|_{L^{2}}^{2}
+4\|(h(u)-P_{1}-\partial_{x}P_{2})(T)\|_{L^{2}}^{2}\\[3pt]
&\quad+2(|\alpha|+|\lambda|)\|(h(u)-P_{1}-\partial_{x}P_{2})(T)\|_{L^{2}}\|v(T)\|_{L^{2}}.
\end{array}
$$
From the proof of Theorem 3.1, we know
$\|h(u)(T)\|_{L^{2}}, \|P_{1}(T)\|_{L^{2}}$ and $\|\partial_{x}P_{2}(T)\|_{L^{2}}$
are uniformly bounded for $T\in I$. Thus,
$\int_{\mathcal{O}(T,\delta)}v_{T}^{2}d Y$ is bounded, which together with (3.13)
implies $\textmd{meas} (\mathcal{O}(T,\delta))$ is finite for all $T\in I$.

For any interior point $\bar{T}$ of $I$, by means of (3.13), we can deduce that there exists a small open interval $J$ such that $\bar{T}\in J$ and
$\mathcal{O}(T,\frac{\delta}{2})\subset \mathcal{O}(\bar{T},\delta)$ for all
$T\in J$. Since $I$ is compact, there exists finitely many points $T_{i}\in I$
and open intervals $J_{i}$ ($i=1,2,\cdots, N$)
such that
$$
I=\cup_{i=1}^{N}(J_{i}\cap I)\quad\mbox{and}\quad
\mathcal{O}(T,\frac{\delta}{2})\subset \mathcal{O}(T_{i},\delta)\quad
\mbox{for all}~ T\in J_{i}\cap I.
$$
From the fact $\{Y\in\mathbb{R};~ v(T, Y)=-\pi\}\subset \mathcal{O}(T,\frac{\delta}{2})$, we know
$$\{Y\in\mathbb{R};~ v(T, Y)=-\pi\}\subset\cup_{i=1}^{N}\mathcal{O}(T_{i},\delta)
\quad \mbox{for~all}~T\in I
$$
and then $\textmd{meas}\{Y\in\mathbb{R};~ v(T, Y)=-\pi\}$ is finite.

Applying the Fubini theorem, we have
$$
\begin{array}{rl}
&\frac{\alpha^{2}}{16}\textmd{meas}\{(T,Y)\in(\mathcal{N}\cap I)\times\mathbb{R};
~ v(T, Y)=-\pi\}
=\frac{\alpha^{2}}{16}\int_{\mathcal{N}\cap I}
\textmd{meas}\{Y\in\mathbb{R};~ v(T, Y)=-\pi\}dT\\[3pt]
&<\int_{\mathcal{N}\cap I}\int_{\{Y\in\mathbb{R};~ v(T, Y)=-\pi\}}
v_{T}^{2}dYdT
=\int_{\mathcal{N}\cap I}\int_{\{Y\in\cup_{i=1}^{N}\mathcal{O}(T_{i},\delta);~ v(T, Y)=-\pi\}}v_{T}^{2}dYdT\\[3pt]
&=\int_{\cup_{i=1}^{N}\mathcal{O}(T_{i},\delta)}
\int_{\{T\in\mathcal{N}\cap I; ~ v(T, Y)=-\pi\}}v_{T}^{2}dTdY
\leq \int_{\cup_{i=1}^{N}\mathcal{O}(T_{i},\delta)}
\int_{\{T\in I; ~ v(T, Y)=-\pi\}}v_{T}^{2}dTdY.
\end{array}
\eqno(3.14)
$$

Now we prove the desired result by using the contradiction
argument. Assume that $\textmd{meas}(\mathcal{N}\cap I)>0$, then (3.14) implies that
$\int_{\cup_{i=1}^{N}\mathcal{O}(T_{i},\delta)}
\int_{\{T\in I; ~ v(T, Y)=-\pi\}}v_{T}^{2}dTdY>0$, which is impossible since $v_{T}(\cdot,Y)=0$ a.e. on $\{T\in I; v(T,Y)=-\pi\}$ due to the locally Lipschitz continuity of the map $T\mapsto v(T,Y)$ at every fixed $Y\in \mathbb{R}$. This completes the proof. \hfill $\Box$

\subsection{Global existence of the weak solution to (3.1)-(3.2)}

In this subsection, we use an inverse transform on the solution of the semi-linear system to construct the solution to the original problem (1.7).

We define $x$ and $t$ as functions of $T$ and $Y$:
$$
\begin{array}{rl}
x(T,Y)
:=\bar{x}(Y)+\int_{0}^{T}(\alpha u(s,Y)+\beta)ds,\quad t=T.
\end{array}
\eqno(3.15)
$$
Thus,
$$
\frac{\partial}{\partial T}x(T,Y)=\alpha u(T,Y)+\beta, \quad
x(0,Y)=\bar{x}(Y),
$$
which means that $x(T,Y)$ is a characteristic.

Next, we show that
$$
u(t,x):=u(T,Y)\quad
\mbox{if}~
x=x(T,Y),~t=T
$$
provides a weak solution of (3.1)-(3.2).\\

\noindent\textbf{Proof of Theorem 1.1.} First, we prove that the function $u=u(t,x)$ is well-defined.
From (3.9), we see that $|u(T,Y)|\leq \mathfrak{C}^{1/2}$ for $T\in [0, \tilde{T}]$.
By (3.15), we have
$$
\bar{x}(Y)-(|\alpha|\mathfrak{C}^{1/2}+|\beta|)T
\leq x(T,Y)\leq \bar{x}(Y)+(|\alpha|\mathfrak{C}^{1/2}+|\beta|)T, \quad \forall T\in [0, \tilde{T}].
$$
From the definition of $Y$ at (2.1), we know
$\lim_{Y\rightarrow \pm \infty}x(T,Y)=\pm \infty$, this yields
the image of the map $(T,Y)\mapsto (T, x(T,Y))$ is the entire half-plane $\mathbb{R}_{+}\times \mathbb{R}$.

We claim that
$$
\begin{array}{rl}
x_{Y}=\xi\cos^{2}\frac{v}{2}\geq 0,
\quad \mbox{for~all}~T\geq0~\mbox{and}~a.e.~Y\in \mathbb{R}.
\end{array}
\eqno(3.16)
$$
Indeed, from (3.1), we deduce that
$$
\begin{array}{rl}
\frac{\partial}{\partial T}(\xi\cos^{2}\frac{v}{2})
=\xi_{T}\cos^{2}\frac{v}{2}
-\xi v_{T}\cos\frac{v}{2}\sin\frac{v}{2}
=\alpha\xi\cos\frac{v}{2}\sin\frac{v}{2}
=\alpha u_{Y}.
\end{array}
$$
By differentiating (3.15) with respect to $T$ and $Y$, we have
$$
\frac{\partial}{\partial T}x_{Y}
=\frac{\partial}{\partial T}(\bar{x}_{Y}+\int_{0}^{T}\alpha u_{Y}ds)
=\alpha u_{Y}=\frac{\partial}{\partial T}(\xi\cos^{2}\frac{v}{2}).
$$
Since the function $x\mapsto 2\arctan u_{0,x}(x)$ is measurable, we see that
the identity (3.16) holds for almost every $Y\in \mathbb{R}$ at $T=0$. Then,
(3.16) remains true for all times $T\geq 0$.

Now we show that $u(t,x)=u(T,x(T,Y))$ is well-defined. We may assume that $Y_{1}< Y_{2}$ but $x(T^{*},Y_{1})=x(T^{*},Y_{2})$, then (3.16) implies that
$$
0=x(T^{*},Y_{1})-x(T^{*},Y_{2})
=\int_{Y_{1}}^{Y_{2}}x_{Y}(T^{*},Y)dY
=\int_{Y_{1}}^{Y_{2}}(\xi\cos^{2}\frac{v}{2})(T^{*},Y)dY.
$$
Then, $\cos\frac{v(T^{*},Y)}{2}\equiv0$ for any $Y\in[Y_{1},Y_{2}]$.
Hence,
$$
\begin{array}{rl}
u(T^{*},Y_{1})-u(T^{*},Y_{2})
=\int_{Y_{1}}^{Y_{2}}u_{Y}(T^{*},Y)dY
=\frac{1}{2}\int_{Y_{1}}^{Y_{2}}
(\xi\sin\frac{v}{2}\cos\frac{v}{2})(T^{*},Y)dY
=0.
\end{array}
$$
This implies that $u(t,x)=u(T,x(T,Y))$ is well-defined for all
$t\geq 0$ and $x\in \mathbb{R}$.

Next, we prove the regularity of $u(t,x)$. Recall that
$u_{x}(t,x)=\tan\frac{v(T,Y)}{2}$
if $x=x(T,Y),~t=T$ and $v(T,Y)\neq -\pi$.
From (3.8), we know that for all
$t\in [0, \tilde{T}]$,
$$
\begin{array}{rl}
\mathfrak{C}
&\geq \int_{\mathbb{R}}
(u^{2}\xi\cos^{2}\frac{v}{2}+\xi\sin^{2}\frac{v}{2})(T,Y)dY\\[3pt]
&\geq
\int_{\{Y\in\mathbb{R}; v(T,Y)\neq -\pi\}}
(u^{2}\xi\cos^{2}\frac{v}{2}+\xi\sin^{2}\frac{v}{2})(T,Y)dY\\[3pt]
&=\int_{\mathbb{R}}(u^{2}+u_{x}^{2})(t,x)dx.
\end{array}
$$
Applying the Sobolev inequality, we know that $u$ as a function of $x$ is H\"{o}lder continuous with exponent $\frac{1}{2}$. On the other hand, similar as the proofs of (3.10)
and (3.11), we can show that $\|\partial_{x}P_{1}(T)\|_{L^{\infty}}, \|P_{2}(T)\|_{L^{\infty}}$ is uniformly bounded for $T\in [0, \tilde{T}]$. Therefore, by the first equation in (3.1), we have
$$
|\frac{d}{dt}u(t,x(t,Y))|=|u_{T}|
\leq|\partial_{x}P_{1}(T,Y)|+|P_{2}(T,Y)|+ |\lambda u(T,Y)|<\infty,
\quad \forall t\in [0, \tilde{T}].
$$
Then, the map $t\mapsto u(t,x(t))$ is locally Lipschitz continuous along every characteristic curve $t\mapsto x(t)$. Therefore, $u=u(t,x)$ is H\"{o}lder continuous on any bounded time interval.

We now prove that the map $t\mapsto u(t)$ is Lipschitz continuous with values
in $L^{2}(\mathbb{R})$ on any bounded time interval. Indeed, we may assume
$t\in [0,\tilde{T}]$ and let $[\tau, \tau+\theta]\subset [0,\tilde{T}]$ be any small interval. For a given point $(\tau, \hat{x})$, we choose the characterstic
$T\mapsto x(T,Y)$ passes through the point $(\tau,\hat{x})$,
i.e. $x(\tau)=\hat{x}$. Since $\|u(t)\|_{L^{\infty}}\leq \mathfrak{C}^{1/2}$
for $t\in [0,\tilde{T}]$, we have
$$
\begin{array}{rl}
&|u(\tau+\theta,\hat{x})-u(\tau,\hat{x})|\\[3pt]
&\leq|u(\tau+\theta,\hat{x})-u(\tau+\theta,x(\tau+\theta,Y))|
+|u(\tau+\theta,x(\tau+\theta,Y))-u(\tau,x(\tau,Y))|\\[3pt]
&\leq \sup_{|y-\hat{x}|\leq (|\alpha|\mathfrak{C}^{1/2}+|\beta|)\theta}|u(\tau+\theta,y)-u(\tau+\theta,\hat{x})|
+\int_{\tau}^{\tau+\theta}|\partial_{x}P_{1}+P_{2}+\lambda u|dt.
\end{array}
$$
Integrating w.r.t. $x$ over $\mathbb{R}$, we have
$$
\begin{array}{rl}
&\int_{\mathbb{R}}|u(\tau+\theta,\hat{x})-u(\tau,\hat{x})|^{2}dx\\[3pt]
&\leq 2\int_{\mathbb{R}}(\int_{\hat{x}-(|\alpha|\mathfrak{C}^{1/2}+|\beta|)\theta}
^{\hat{x}+(|\alpha|\mathfrak{C}^{1/2}+|\beta|)\theta}
|u_{x}(\tau+\theta,y)|dy)^{2}dx
+2\int_{\mathbb{R}}(\int_{\tau}^{\tau+\theta}
|\partial_{x}P_{1}+P_{2}+\lambda u|dt)^{2}\xi(\tau,Y)dY\\[3pt]
&\leq
4(|\alpha|\mathfrak{C}^{1/2}+|\beta|)\theta\int_{\mathbb{R}}
\int_{\hat{x}-(|\alpha|\mathfrak{C}^{1/2}+|\beta|)\theta}^{\hat{x}+(|\alpha|\mathfrak{C}^{1/2}+|\beta|)\theta}
|u_{x}(\tau+\theta,y)|^{2}dydx\\[3pt]
&\quad+2\theta\int_{\mathbb{R}}(\int_{\tau}^{\tau+\theta}
|\partial_{x}P_{1}+P_{2}+\lambda u|^{2}dt)\|\xi(\tau)\|_{L^{\infty}}dY\\[3pt]
&\leq 8(|\alpha|\mathfrak{C}^{1/2}+|\beta|)^{2}\theta^{2}\|u_{x}(\tau+\theta)\|_{L^{2}}^{2}
+2\theta\|\xi(\tau)\|_{L^{\infty}}\int_{\tau}^{\tau+\theta}
\|(\partial_{x}P_{1}+P_{2}+\lambda u)(t)\|_{L^{2}}^{2}dt.
\end{array}
$$
This implies the locally Lipschitz continuity of the map
$t\mapsto u(t)$, in terms of the $x$-variable.

Next, we prove that the identities (1.10) and (1.11) hold for any test function $\phi(t,x)\in C_{c}^{1}(\Gamma)$, which imply that the function $u$ provides a weak solution of (1.8).
We denote
$$
\Gamma=[0,\infty)\times \mathbb{R}, \quad
\overline{\Gamma}=\Gamma\cap\{(T,Y); v(T,Y)\neq -\pi\},
$$
and $\phi(T,Y)=\phi(T,x(T,Y))$ as explained in Section 2. In view of (2.3), (3.7), (3.16) and Lemma 3.3, a direct computation shows that
$$
\begin{array}{rl}
0
&=\int\int_{\Gamma}
\{u_{YT}\phi+[h(u)\cos^{2}\frac{v}{2}
+\frac{\alpha}{2}\sin^{2}\frac{v}{2}
-(P_{1}+\partial_{x}P_{2})\cos^{2}\frac{v}{2}
-\frac{1}{2}\lambda\sin v]\xi\phi\}(T,Y)dYdT\\[3pt]
&=\int\int_{\Gamma}
\{-u_{Y}\phi_{T}+[h(u)\cos^{2}\frac{v}{2}
+\frac{\alpha}{2}\sin^{2}\frac{v}{2}
-(P_{1}+\partial_{x}P_{2})\cos^{2}\frac{v}{2}
-\frac{1}{2}\lambda\sin v]\xi\phi\}(T,Y)dYdT\\[3pt]
&=\int\int_{\overline{\Gamma}}
\{-u_{Y}\phi_{T}+[h(u)\cos^{2}\frac{v}{2}
+\frac{\alpha}{2}\sin^{2}\frac{v}{2}
-(P_{1}+\partial_{x}P_{2})\cos^{2}\frac{v}{2}
-\frac{1}{2}\lambda\sin v]\xi\phi\}(T,Y)dYdT\\[3pt]
&=\int\int_{\Gamma}
\{-u_{x}[\phi_{t}+(\alpha u+\beta)\phi_{x}]+[P_{1}+\partial_{x}P_{2}-h(u)-\frac{\alpha}{2}u_{x}^{2}+\lambda u_{x}]\phi\}(t,x)dxdt,
\end{array}
$$
which implies (1.10) holds. Now, we introduce the Radon measures
$\{\mu_{(t)}, t\in \mathbb{R}_{+}\}$ as follows
$$
\begin{array}{rl}
\mu_{(t)}(\mathfrak{M})
=e^{2\lambda t}\int_{\{Y\in \mathbb{R}; x(t,Y)\in \mathfrak{M}\}}
(\xi\sin^{2}\frac{v}{2})(t,Y)dY
\end{array}
$$
for any Lebesgue measurable set $\mathfrak{M}\subset\mathbb{R}$.
For every $t\notin \mathcal{N}$, the absolutely continuous part of
$\mu_{(t)}$ w.r.t. Lebesgue measure has density $e^{2\lambda t}u_{x}^{2}(t,\cdot)$
by (3.16). It follows from (3.1) and Lemma 3.3 that for any test function $\phi(t,x)\in C_{c}^{1}(\Gamma)$,
$$
\begin{array}{rl}
&-\int_{\mathbb{R}_{+}}
\{\int[\phi_{t}+(\alpha u+\beta)\phi_{x}]d\mu_{(t)}\}dt
=-\int\int_{\Gamma}\phi_{T}e^{2\lambda T}\xi\sin^{2}\frac{v}{2}dYdT\\[3pt]
&=\int\int_{\Gamma}\phi(e^{2\lambda T}\xi\sin^{2}\frac{v}{2})_{T}dYdT
=\int\int_{\Gamma}\phi
[2\lambda e^{2\lambda T}\xi\sin^{2}\frac{v}{2}
+e^{2\lambda T}(\xi\sin^{2}\frac{v}{2})_{T}]dYdT\\[3pt]
&=\int\int_{\Gamma}\phi
[2\lambda e^{2\lambda T}\xi\sin^{2}\frac{v}{2}
+e^{2\lambda T}\xi_{T}\sin^{2}\frac{v}{2}
+e^{2\lambda T}\xi v_{T}\sin\frac{v}{2}\cos\frac{v}{2}]dYdT\\[3pt]
&=\int\int_{\Gamma}2e^{2\lambda T}(h(u)-P_{1}-\partial_{x}P_{2})\xi\phi\sin\frac{v}{2}\cos\frac{v}{2} dYdT\\[3pt]
&=\int\int_{\overline{\Gamma}}2e^{2\lambda T}(h(u)-P_{1}-\partial_{x}P_{2})\xi\phi\sin\frac{v}{2}\cos\frac{v}{2} dYdT\\[3pt]
&=\int\int_{\Gamma}2e^{2\lambda t}(h(u)-P_{1}-\partial_{x}P_{2})u_{x}\phi dxdt.
\end{array}
$$
Thus, (1.11) holds.
Following the arguments in \cite{brco07}, we can easily obtain the continuous dependence result. \hfill $\Box$

\section{Uniqueness of the global weak solution}

In this section, motivated by the work \cite{bcz15}, we prove the global weak solution satisfying the initial data in $L^{2}(\mathbb{R})$ together with (1.10) and (1.11) is unique. By introducing a new energy variable $\eta$, we first prove that $u(t,x)$ satisfies a semi-linear system under new independent variables $(t,\eta)$. Then by using the uniqueness of the solution to the new semi-linear system, we obtain the uniqueness of global weak solution of (1.7).

For any time $t\in \mathbb{R}_{+}$ and $\eta\in \mathbb{R}$, we define $x(t,\eta)$ to be the unique point $\bar{x}$ such that
$$
\begin{array}{rl}
\bar{x}+\mu_{(t)}\{(-\infty,\bar{x})\}
\leq \eta\leq \bar{x}+\mu_{(t)}\{(-\infty,\bar{x}]\}.
\end{array}
\eqno(4.1)
$$
Hence,
$$
\begin{array}{rl}
\eta=x(t, \eta)+\mu_{(t)}\{(-\infty,x(t, \eta))\}
+\theta\cdot \mu_{(t)}\{x(t, \eta)\}
\end{array}
\eqno(4.2)
$$
for some $\theta\in[0,1]$. Note that at every time where $\mu_{(t)}$ is absolutely continuous with density $e^{2\lambda t}u_{x}^{2}$ w.r.t. Lebesgue measure, the above definition gives
$$
\begin{array}{rl}
\eta
=x(t, \eta)+\mu_{(t)}\{(-\infty,x(t, \eta))\}
=
x(t,\eta)+e^{2\lambda t}\int_{-\infty}^{x(t,\eta)}u_{x}^{2}(t,z)dz.
\end{array}
\eqno(4.3)
$$
Now we study the Lipschitz continuity of $x$ and $u$ as functions of $t, \eta$.\\

\noindent\textbf{Lemma 4.1.}
Let $u=u(t,x)$ be the weak solution of (1.7) satifying (1.10)-(1.11).
Then,\\
$(i)$~for every fixed $t\geq0$, $\eta\mapsto x(t,\eta)$ and $\eta\mapsto u(t,\eta):=u(t,x(t,\eta))$
are Lipschitz continuous with constants $1$
and $\frac{1}{2}\max\{1,e^{-2\lambda t}\}$, respectively;\\
$(ii)$~the map $t\mapsto x(t,\eta)$ is locally Lipschitz continuous with a constant depending on $\|u_{0}\|_{H^{1}}$ and time interval.\\

\noindent\textbf{Proof.}
$(i)$ For any fixed time $t\geq 0$, the map
$x\mapsto \eta(t,x)$ is right continuous and strictly increasing. Hence it has a well-defined,
continuous and nondecreasing inverse $\eta\mapsto x(t, \eta)$.
If $\eta_{1}<\eta_{2}$, then
$$
\begin{array}{rl}
x(t,\eta_{2})-x(t,\eta_{1})
+\mu_{(t)}\{(x(t,\eta_{1}), x(t,\eta_{2}))\}
\leq \eta_{2}-\eta_{1}.
\end{array}
\eqno(4.4)
$$
This implies
$$
x(t,\eta_{2})-x(t,\eta_{1})
\leq \eta_{2}-\eta_{1},
$$
showing that the map $\eta\mapsto x(t,\eta)$ is Lipschitz continuous with constant $1$.

To prove the Lipschitz continuity of the map $\eta\mapsto u(t,\eta)$,
we assume $\eta_{1}<\eta_{2}$. By (4.4) it follows
$$
\begin{array}{rl}
|u(t,x(t,\eta_{2}))-u(t,x(t,\eta_{1}))|
&\leq\int_{x(t,\eta_{1})}^{x(t,\eta_{2})}|u_{x}|dx
\leq \int_{x(t,\eta_{1})}^{x(t,\eta_{2})}\frac{1}{2}(e^{-2\lambda t}+e^{2\lambda t}u_{x}^{2})dx\\[3pt]
&\leq \frac{1}{2}\max\{1,e^{-2\lambda t}\}[x(t,\eta_{2})-x(t,\eta_{1})
+\mu_{(t)}\{(x(t,\eta_{1}), x(t,\eta_{2}))\}]\\[3pt]
&\leq \frac{1}{2}\max\{1,e^{-2\lambda t}\}(\eta_{2}-\eta_{1}).
\end{array}
\eqno(4.5)
$$

$(ii)$ We prove the Lipschitz continuity of the map
$t\mapsto x(t,\eta)$ on $[0,\tilde{T}]$ with any fixed $\tilde{T}>0$.
Recall that the family of measure $\mu_{(t)}$ satifies the balance law (1.11), where for each
$t\in [0,\tilde{T}]$ the source term $2e^{2\lambda t}(h(u)-P_{1}-\partial_{x}P_{2})u_{x}$
satisfies
$$
\|2e^{2\lambda t}[(h(u)-P_{1}-\partial_{x}P_{2})u_{x}](t)\|_{L^{1}}
\leq 2e^{2\lambda \tilde{T}}\|(h(u)-P_{1}-\partial_{x}P_{2})(t)\|_{L^{2}}\|u_{x}(t)\|_{L^{2}}
\leq C_{0}
$$
for some constant $C_{0}$ depending on $\|u\|_{H^{1}}$ and $\tilde{T}$.
For any $0\leq\tau<t\leq\tilde{T}$,
$$
\begin{array}{rl}
\mu_{(t)}\{(-\infty,y-C_{\infty}(t-\tau))\}
&\leq \mu_{(\tau)}\{(-\infty,y)\}
+\int_{\tau}^{t}\|2e^{2\lambda t}[(h(u)-P_{1}-\partial_{x}P_{2})u_{x}](t)\|_{L^{1}}dt\\[3pt]
&\leq \mu_{(\tau)}\{(-\infty,y)\}
+C_{0}(t-\tau),
\end{array}
$$
where $y:=x(\tau,\eta)$ and $C_{\infty}:=\frac{|\alpha|}{\sqrt{2}}\|u\|_{H^{1}}+|\beta|$.

Let $y^{-}(t):=y-(C_{\infty}+C_{0})(t-\tau)$. Then, we have
$$
\begin{array}{rl}
y^{-}(t)+\mu_{(t)}\{(-\infty,y^{-}(t))\}
&\leq y-(C_{\infty}+C_{0})(t-\tau)
+\mu_{(\tau)}\{(-\infty,y)\}+C_{0}(t-\tau)\\[3pt]
&\leq y+\mu_{(\tau)}\{(-\infty,y)\}
\leq \eta.
\end{array}
$$
This implies that $x(t, \eta)\geq y^{-}(t)$ for all $0\leq\tau<t\leq\tilde{T}$.
Similarly, we can obtain that
$x(t, \eta)\leq y^{+}(t):=y+(C_{\infty}+C_{0})(t-\tau)$. This completes
the proof of locally Lipschitz continuity of the mapping $t\mapsto x(t,\eta)$.  \hfill $\Box$\\

\noindent\textbf{Lemma 4.2.}
Let $u=u(t,x)$ be the weak solution of (1.7) satisfying (1.10)-(1.11). Then, for any $x_{0}\in \mathbb{R}$,
there exists a unique locally Lipschitz continuous map $t\mapsto x(t)$ which satisfies
$$
\begin{array}{rl}
\frac{d}{dt}x(t)=\alpha u(t,x(t))+\beta, \quad x(0)=x_{0}
\end{array}
\eqno(4.6)
$$
and
$$
\begin{array}{rl}
&\frac{d}{dt}\left(\mu_{(t)}\{(-\infty,x(t))\}
+\theta(t,x_{0})\cdot \mu_{(t)}\{x(t)\}\right)
=\int_{-\infty}^{x(t)}2e^{2\lambda t}[(h(u)-P_{1}-\partial_{x}P_{2})u_{x}](t,z)dz,\\[3pt]
&x(0)=x_{0},
\end{array}
\eqno(4.7)
$$
for some function $\theta\in [0,1]$ and almost every time $t\geq 0$. Furthermore, for any $0\leq \tau_{0}<\tau<\infty$, we have
$$
\begin{array}{rl}
u(t,x(\tau))-u(\tau,x(\tau_{0}))
=-\int_{\tau_{0}}^{\tau}(\partial_{x}P_{1}+P_{2}+\lambda u)(s,x(s))ds.
\end{array}
\eqno(4.8)
$$

\noindent\textbf{Proof.} Firstly, by the adapted coordinates $(t,\eta)$, we write the characteristic starting at $x_{0}$ in the form $t\mapsto x(t)=x(t,\eta(t))$, where
$\eta(\cdot)$ is a map to be determined. Summing up (4.6) and (4.7) and integrating w.r.t. time $t$, we get
$$
\begin{array}{rl}
\eta(t)
&=x(t)+\mu_{(t)}\{(-\infty,x(t))\}
+\theta(t)\cdot \mu_{(t)}\{x(t)\}\\[3pt]
&=\bar{\eta}+\int_{0}^{t}\{\beta+\int_{-\infty}^{x(s)}
[\alpha u_{x}+2e^{2\lambda s}(h(u)-P_{1}-\partial_{x}P_{2})u_{x}](s,z)dz\}ds,
\end{array}
\eqno(4.9)
$$
where
$\bar{\eta}=x_{0}+\int_{-\infty}^{x_{0}}u_{0,x}^{2}dx$.
Introducing the function
$$
\begin{array}{rl}
G(t,\eta)
:=\beta+\int_{-\infty}^{x(t,\eta)}
[\alpha u_{x}+2e^{2\lambda t}(h(u)-P_{1}-\partial_{x}P_{2})u_{x}](t,z)dz,
\end{array}
\eqno(4.10)
$$
then (4.9) can be written as
$$
\begin{array}{rl}
\eta(t)
=\bar{\eta}+\int_{0}^{t}G(s,\eta(s))ds.
\end{array}
\eqno(4.11)
$$

For each $t\in [0,\tilde{T}]$ with any fixed $\tilde{T}>0$, since the maps $x\mapsto u(t,x)$ and $x\mapsto P_{1}(t,x)$
and $x\mapsto \partial_{x}P_{2}(t,x)$ are both in $H^{1}(\mathbb{R})$, the function
$\eta\mapsto G(t,\eta)$ defined in (4.10) is uniformly bounded and absolutely continuous. Moreover, from (4.2), (4.3) and (4.10), we know that if $\mu_{(t)}\{x(t,\eta)\}\neq0$ with $\eta\in(x(t,\eta)+\mu_{(t)}\{(-\infty,x(t,\eta))\},
x(t,\eta)+\mu_{(t)}\{(-\infty,x(t,\eta)]\})$, then $G_{\eta}(t,\eta)=0$,
and if $\mu_{(t)}\{x(t,\eta)\}=0$, then
$$
\begin{array}{rl}
|G_{\eta}(t,\eta)|
&=\left|[\alpha u_{x}+2e^{2\lambda t}(h(u)-P_{1}-\partial_{x}P_{2})u_{x}]x_{\eta}\right|\\[3pt]
&=\Big|\dfrac{\alpha u_{x}+2e^{2\lambda t}(h(u)-P_{1}-\partial_{x}P_{2})u_{x}}{1+e^{2\lambda t}u_{x}^{2}}\Big|\\[3pt]
&\leq \dfrac{\frac{|\alpha|}{2}(e^{-2\lambda t}+e^{2\lambda t}u_{x}^{2})
+e^{\lambda t}(1+e^{2\lambda t}u_{x}^{2})
\|(h(u)-P_{1}-\partial_{x}P_{2})(t)\|_{L^{\infty}}}{1+e^{2\lambda t}u_{x}^{2}}\\[3pt]
&\leq \frac{|\alpha|}{2}\max\{1, e^{-2\lambda t}\}
+e^{\lambda t}\|(h(u)-P_{1}-\partial_{x}P_{2})(t)\|_{L^{\infty}}
\leq C
\end{array}
$$
for some constant $C$ depending on $\|u\|_{H^{1}}$ and $\tilde{T}$.
Hence, the function $G$ is Lipschitz continuous w.r.t. $\eta$ for $t\in [0,\tilde{T}]$. We can apply the ODE's theory
in the Banach space of all continuous functions $\eta: [0,\tilde{T}]\rightarrow \mathbb{R}$ with
weighted norm $\|\eta\|_{*}:=\sup_{t\in [0,\tilde{T}]}e^{-2Ct}|\eta(t)|$.
Let $(\varphi \eta)(t):=\bar{\eta}+\int_{0}^{t}G(s, \eta(s))ds$. Assume
$\|\eta_{1}-\eta_{2}\|_{*}=\delta>0$, that is,
$|\eta_{1}(s)-\eta_{2}(s)|\leq \delta e^{2Cs}$ for all $s\in [0,\tilde{T}]$.
By the Lipschitz continuity of $G$,
$$
|(\varphi\eta_{1})(t)-(\varphi\eta_{2})(t)|
\leq C\int_{0}^{t}|\eta_{1}(s)-\eta_{2}(s)|ds
\leq \frac{\delta}{2}e^{2Ct},
$$
which implies $\|\varphi\eta_{1}-\varphi\eta_{2}\|_{*}\leq \frac{\delta}{2}$. Thus, the map
$(\varphi \eta)(t)$ is a strict contraction for $t\in [0,\tilde{T}]$. By the contraction mapping principle, the integral equation (4.11) has a unique solution defined on $[0,\tilde{T}]$.

By the previous construction, the map $t\mapsto x(t)=x(t,\eta(t))$ provides the unique solution of (4.9) with $t\in [0,\tilde{T}]$.
The arbitrariness of $\tilde{T}$ and the uniqueness of solution imply that
the solution is defined for all $t\in \mathbb{R}_{+}$.
Being the composition of two Lipschitz functions, the
map $t\mapsto x(t,\eta(t))$ is locally Lipschitz continuous. To prove that it satisfies the ODE for
characteristics (4.6), it suffices to show that (4.6) holds at each time $\tau>0$ satisfying that $x(\cdot)$ is differentiable at $t=\tau$ and the measure $\mu_{(\tau)}$ is absolutely continuous.

Assume, on the contrary, that $\frac{d}{dt}x(\tau_{0})\neq \alpha u(\tau_{0},x(\tau_{0}))+\beta$.
Let
$$
\frac{d}{dt}x(\tau_{0})= \alpha u(\tau_{0},x(\tau_{0}))+\beta+2\varepsilon_{0}
$$
for some $\varepsilon_{0}>0$. The case $\varepsilon_{0}<0$ is entirely similar.
To derive a contradiction we observe that, for all $\tau\in (\tau_{0},\tau_{0}+\delta]$
with $0<\delta\ll 1$, one has
$$
\begin{array}{rl}
x^{+}(\tau):=
x(\tau_{0})+(\tau-\tau_{0})[\alpha u(\tau_{0},x(\tau_{0}))+\beta+\varepsilon_{0}]<x(\tau).
\end{array}
\eqno(4.12)
$$
We also see that if $\phi$ is Lipschitz continuous with compact support then
(1.12) is still true. For any $\epsilon>0$ small, we can still use the following test functions used in \cite{bcz15}:
$$
\begin{array}{rl}
\rho^{\epsilon}(t,x)=
\left\{\begin{array}{l}
0, \quad \mbox{if}~x\leq -\epsilon^{-1}, \\[3pt]
x+\epsilon^{-1}, \quad \mbox{if}~-\epsilon^{-1}\leq x\leq 1-\epsilon^{-1}, \\[3pt]
1, \quad \mbox{if}~1-\epsilon^{-1}\leq x\leq x^{+}(t), \\[3pt]
1-\epsilon^{-1}(x-x^{+}(t)), \quad \mbox{if}~x^{+}(t)\leq x\leq x^{+}(t)+\epsilon, \\[3pt]
0, \quad \mbox{if}~x\geq x^{+}(t)+\epsilon,
\end{array}
\right.
\end{array}
$$

$$
\begin{array}{rl}
\chi^{\epsilon}(t)=
\left\{\begin{array}{l}
0, \quad \mbox{if}~t\leq \tau_{0}-\epsilon, \\[3pt]
\epsilon^{-1}(t-\tau_{0}+\epsilon), \quad \mbox{if}~\tau_{0}-\epsilon\leq t\leq \tau_{0}, \\[3pt]
1, \quad \mbox{if}~\tau_{0}\leq t\leq \tau, \\[3pt]
1-\epsilon^{-1}(t-\tau), \quad \mbox{if}~\tau\leq t< \tau+\epsilon, \\[3pt]
0, \quad \mbox{if}~t\geq \tau+\epsilon,
\end{array}
\right.
\end{array}
\eqno(4.13)
$$
where $x$ is spatial variable and $x^{+}(t)$ is defined in (4.12).
Define
$$
\phi^{\epsilon}(t,x):=
\min\{\rho^{\epsilon}(t,x), \chi^{\epsilon}(t)\}.
$$
Using $\phi^{\epsilon}$ as test function in (1.11) we obtain
$$
\begin{array}{rl}
\int_{\mathbb{R}^{+}}\left(\int[\phi^{\epsilon}_{t}+(\alpha u+\beta)\phi^{\epsilon}_{x}]d\mu_{(t)}
+\int 2e^{2\lambda t}(h(u)-P_{1}-\partial_{x}P_{2})u_{x}\phi^{\epsilon} dx\right)dt
=0.
\end{array}
\eqno(4.14)
$$
Note that $|\tau-\tau_{0}|\ll 1$. For $s\in[\tau_{0}+\epsilon, \tau-\epsilon]$ and
$z\in(x^{+}(s)-\epsilon,x^{+}(s))\cup(x^{+}(s), x^{+}(s)+\epsilon)$, one has
$$
\begin{array}{rl}
0=\phi^{\epsilon}_{t}(s,z)+(\alpha u(\tau_{0},x(\tau_{0}))+\beta+\varepsilon_{0})\phi^{\epsilon}_{x}(s,z)
\leq [\phi^{\epsilon}_{t}+(\alpha u(s,x)+\beta)\phi^{\epsilon}_{x}](s,z),
\end{array}
\eqno(4.15)
$$
since $\alpha u(s,z)+\beta<\alpha u(\tau_{0},x(\tau_{0}))+\beta+\varepsilon_{0}$ and
$\phi^{\epsilon}_{x}\leq 0$. Thus, by (4.15) we have
$$
\lim_{\epsilon\rightarrow 0}
\int_{\tau_{0}}^{\tau}\int_{(x^{+}(s)-\epsilon,x^{+}(s)+\epsilon)}
[\phi^{\epsilon}_{t}+(\alpha u+\beta)\phi^{\epsilon}_{x}]d\mu_{(s)}ds\geq 0.
$$
Since the family of measures $\mu_{(t)}$ depends continuously on $t$ in the topology of
weak convergence, taking the limit of (4.14) as $\epsilon\rightarrow 0$, it is thereby inferred that
$$
\begin{array}{rl}
&\mu_{(\tau)}\{(-\infty,x^{+}(\tau)]\}\\[3pt]
&=\mu_{(\tau_{0})}\{(-\infty,x^{+}(\tau_{0})]\}
+\lim_{\epsilon\rightarrow 0}
\int_{\tau_{0}}^{\tau}\int_{(x^{+}(s)-\epsilon,x^{+}(s)+\epsilon)}
[\phi^{\epsilon}_{t}+(\alpha u+\beta)\phi^{\epsilon}_{x}]d\mu_{(s)}ds\\[3pt]
&\quad+\int_{\tau_{0}}^{\tau}\int_{-\infty}^{x^{+}(s)}2e^{2\lambda s}
\left[(h(u)-P_{1}-\partial_{x}P_{2})u_{x}\right](s,z)dzds\\[3pt]
&\geq \mu_{(\tau_{0})}\{(-\infty,x^{+}(\tau_{0})]\}
+\int_{\tau_{0}}^{\tau}\int_{-\infty}^{x^{+}(s)}2e^{2\lambda s}
\left[(h(u)-P_{1}-\partial_{x}P_{2})u_{x}\right](s,z)dzds\\[3pt]
&=\mu_{(\tau_{0})}\{(-\infty,x^{+}(\tau_{0})]\}
+\int_{\tau_{0}}^{\tau}\int_{-\infty}^{x(s)}2e^{2\lambda s}
\left[(h(u)-P_{1}-\partial_{x}P_{2})u_{x}\right](s,z)dzds
+o_{1}(\tau-\tau_{0}),
\end{array}
$$
where the last term is a higher order infinitesimal, that is,
$\frac{o_{1}(\tau-\tau_{0})}{\tau-\tau_{0}}\rightarrow 0$ as $\tau\rightarrow \tau_{0}$.
Indeed,
$$
\begin{array}{rl}
|o_{1}(\tau-\tau_{0})|
&=|\int_{\tau_{0}}^{\tau}\int_{x^{+}(s)}^{x(s)}2e^{2\lambda s}
\left[(h(u)-P_{1}-\partial_{x}P_{2})u_{x}\right](s,z)dzds|\\[3pt]
&\leq \|2(h(u)-P_{1}-\partial_{x}P_{2})(s)\|_{L^{\infty}}
\int_{\tau_{0}}^{\tau}e^{2\lambda s}|x(s)-x^{+}(s)|^{\frac{1}{2}}
\|u_{x}(s)\|_{L^{2}}ds\\[3pt]
&\leq Ce^{2\lambda(\tau_{0}+1)}(\tau-\tau_{0})^{\frac{3}{2}},
\end{array}
$$
where $C$ is a constant depending on $\|u_{0}\|_{H^{1}}$ and $\tau_{0}$.
For $\tau$ sufficiently close to $\tau_{0}$, we have
$$
\begin{array}{rl}
\eta(\tau)
&=x(\tau)+\mu_{(\tau)}\{(-\infty,x(\tau))\}
+\theta(\tau)\cdot\mu_{(\tau)}\{x(\tau)\}\\[3pt]
&\geq x(\tau_{0})+(\tau-\tau_{0})[\alpha u(\tau_{0},x(\tau_{0}))+\beta+\varepsilon_{0}]
+\mu_{(\tau)}\{(-\infty,x^{+}(\tau)]\}\\[3pt]
&\geq x(\tau_{0})+(\tau-\tau_{0})[\alpha u(\tau_{0},x(\tau_{0}))+\beta+\varepsilon_{0}]
+\mu_{(\tau_{0})}\{(-\infty,x^{+}(\tau_{0})]\}\\[3pt]
&\quad+\int_{\tau_{0}}^{\tau}\int_{-\infty}^{x(s)}2e^{2\lambda s}
\left[(h(u)-P_{1}-\partial_{x}P_{2})u_{x}\right](s,z)dzds
+o_{1}(\tau-\tau_{0}).
\end{array}
\eqno(4.16)
$$

On the other hand, from (4.10) and (4.11), a linear approximation yields
$$
\begin{array}{rl}
\eta(\tau)=\eta(\tau_{0})
+(\tau-\tau_{0})\Big[\beta+\alpha u(\tau_{0},x(\tau_{0}))
+\int_{-\infty}^{x(\tau_{0})}2e^{2\lambda \tau_{0}}[(h(u)-P_{1}-\partial_{x}P_{2})u_{x}](\tau_{0},z)dz\Big]+o_{2}(\tau-\tau_{0})
\end{array}
\eqno(4.17)
$$
with $\frac{o_{2}(\tau-\tau_{0})}{\tau-\tau_{0}}\rightarrow 0$ as $\tau\rightarrow \tau_{0}$.

Combining (4.16) and (4.17), we find
$$
\begin{array}{rl}
&\eta(\tau_{0})
+(\tau-\tau_{0})\Big[\beta+\alpha u(\tau_{0},x(\tau_{0}))
+\int_{-\infty}^{x(\tau_{0})}2e^{2\lambda \tau_{0}}[(h(u)-P_{1}-\partial_{x}P_{2})u_{x}](\tau_{0},z)dz\Big]+o_{2}(\tau-\tau_{0})\\[3pt]
&\geq x(\tau_{0})+(\tau-\tau_{0})[\alpha u(\tau_{0},x(\tau_{0}))+\beta+\varepsilon_{0}]
+\mu_{(\tau_{0})}\{(-\infty,x^{+}(\tau_{0})]\}\\[3pt]
&\quad+\int_{\tau_{0}}^{\tau}\int_{-\infty}^{x(s)}2e^{2\lambda s}
\left[(h(u)-P_{1}-\partial_{x}P_{2})u_{x}\right](s,z)dzds
+o_{1}(\tau-\tau_{0}).
\end{array}
$$
Subtracting common terms, dividing both sides by $\tau-\tau_{0}$ and letting
$\tau\rightarrow \tau_{0}$, we get $\varepsilon_{0}\leq 0$, which is a contradiction.
Therefore, (4.6) must hold.

Next, we prove (4.8) holds for all $0\leq \tau_{0}<\tau<\tilde{T}$ with any fixed $\tilde{T}>0$. By (1.10), for every test function $\phi\in C_{c}^{\infty}$, one has
$$
\int_{0}^{\infty}\int[u\phi_{t}
+(\frac{\alpha}{2}u^{2}+\beta u)\phi_{x}
-(\partial_{x}P_{1}+P_{2}+\lambda u)\phi]dxdt
+\int u_{0}(x)\phi(0,x)dx=0.
$$
Given any $\varphi\in C_{c}^{\infty}$,
we let $\phi=\varphi_{x}$.
Since the map $x\mapsto u(t,x)$ is absolutely continuous, we can integrate by parts w.r.t. $x$
and obtain
$$
\begin{array}{rl}
\int_{0}^{\infty}\int[u_{x}\varphi_{t}
+(\alpha u+\beta)u_{x}\varphi_{x}
+(\partial_{x}P_{1}+P_{2}+\lambda u)\varphi_{x}]dxdt
+\int \partial_{x}u_{0}(x)\varphi(0,x)dx=0.
\end{array}
\eqno(4.18)
$$
By an approximation argument we know that for any test function $\varphi$ which is Lipschitz
continuous with compact support, the identity (4.18) remains valid. For any $\epsilon>0$ sufficiently
small, we consider the functions
$$
\begin{array}{rl}
\varrho^{\epsilon}(t,x)=
\left\{\begin{array}{l}
0, \quad \mbox{if}~x\leq -\epsilon^{-1}, \\[3pt]
x+\epsilon^{-1}, \quad \mbox{if}~-\epsilon^{-1}\leq x\leq 1-\epsilon^{-1}, \\[3pt]
1, \quad \mbox{if}~1-\epsilon^{-1}\leq x\leq x(t), \\[3pt]
1-\epsilon^{-1}(x-x(t)), \quad \mbox{if}~x(t)\leq x\leq x(t)+\epsilon, \\[3pt]
0, \quad \mbox{if}~x\geq x(t)+\epsilon
\end{array}
\right.
\end{array}
$$
and
$$
\psi^{\epsilon}(t,x)
:=\min\{\varrho^{\epsilon}(t,x),\chi^{\epsilon}(t)\},
$$
where $x$ is spatial variable, $x(t)=x(t,\eta(t))$ and
$\chi^{\epsilon}(t)$ is defined in (4.13).
Take $\varphi=\psi^{\epsilon}$ in (4.18) and let $\epsilon\rightarrow 0$.
Observing that the function $(\partial_{x}P_{1}+P_{2}+\lambda u)$ is
continuous, we obtain
$$
\begin{array}{rl}
\int_{-\infty}^{x(\tau)}u_{x}(\tau,z)dz
&=\int_{-\infty}^{x(\tau_{0})}u_{x}(\tau_{0},z)dz
-\int_{\tau_{0}}^{\tau}(\partial_{x}P_{1}+P_{2}+\lambda u)(s,x(s))ds\\[3pt]
&\quad+\lim_{\epsilon\rightarrow 0}
\int_{\tau_{0}-\epsilon}^{\tau+\epsilon}\int_{x(s)}^{x(s)+\epsilon}
u_{x}[\psi^{\epsilon}_{t}
+(\alpha u+\beta)\psi^{\epsilon}_{x}](s,z)dzds.
\end{array}
\eqno(4.19)
$$
To completes the proof it suffices to show that the last term on the right
hand side of (4.19) vanishes for all $0\leq \tau_{0}<\tau<\tilde{T}$.  Since $u_{x}\in L^{2}$, the Cauchy inequality yields
$$
\begin{array}{rl}
|\int_{\tau_{0}}^{\tau}\int_{x(s)}^{x(s)+\epsilon}
u_{x}[\psi^{\epsilon}_{t}
+(\alpha u+\beta)\psi^{\epsilon}_{x}](s,z)dzds|
\leq \int_{\tau_{0}}^{\tau}(\int_{x(s)}^{x(s)+\epsilon}u_{x}^{2}dz)^{\frac{1}{2}}
(\int_{x(s)}^{x(s)+\epsilon}[\psi^{\epsilon}_{t}
+(\alpha u+\beta)\psi^{\epsilon}_{x}]^{2}dz)^{\frac{1}{2}}ds.
\end{array}
\eqno(4.20)
$$
For each $\epsilon>0$, consider the function
$$
\varsigma_{\epsilon}(s)
:=(\sup_{y\in \mathbb{R}}\int_{y}^{y+\epsilon}u_{x}^{2}(s,z)dz)^{\frac{1}{2}}.
$$
Observe that all functions $\varsigma_{\epsilon}$ are uniformly
bounded for $s\in [0,\tilde{T})$. Furthermore, we have $\varsigma_{\epsilon}(s)\rightarrow 0$ pointwise at a.e. $s\in [0,\tilde{T})$ as $\epsilon\rightarrow 0$.
Therefore, in view of the dominated convergence theorem, we have
$$
\begin{array}{rl}
\lim_{\epsilon\rightarrow 0}\int_{\tau_{0}}^{\tau}
(\int_{x(s)}^{x(s)+\epsilon}u_{x}^{2}(s,z)dz)^{\frac{1}{2}}ds
\leq \lim_{\epsilon\rightarrow 0}\int_{\tau_{0}}^{\tau}\varsigma_{\epsilon}(s)ds=0.
\end{array}
\eqno(4.21)
$$
For every time $s\in[\tau_{0}, \tau]$ and $x(s)<z<x(s)+\epsilon$, by the definition of $\psi^{\epsilon}$, we have
$$
\begin{array}{rl}
\psi^{\epsilon}_{x}(s,z)=\epsilon^{-1},\quad
\psi^{\epsilon}_{t}(s,z)
+(\alpha u(s,x(s))+\beta)\psi^{\epsilon}_{x}(s,z)=0.
\end{array}
$$
This implies
$$
\begin{array}{rl}
&\int_{x(s)}^{x(s)+\epsilon}
|\psi^{\epsilon}_{t}(s,z)
+(\alpha u(s,z)+\beta)\psi^{\epsilon}_{x}(s,z)|^{2}dz\\[3pt]
&=\frac{\alpha^{2}}{\epsilon^{2}}\int_{x(s)}^{x(s)+\epsilon}
|u(s,z)-u(s,x(s))|^{2}dz\\[3pt]
&\leq \frac{\alpha^{2}}{\epsilon}(\max_{x(s)\leq z\leq x(s)+\epsilon}|u(s,z)-u(s,x(s))|)^{2}\\[3pt]
&\leq \frac{\alpha^{2}}{\epsilon}
(\int_{x(s)}^{x(s)+\epsilon}|u_{x}(s,z)|dz)^{2}
\leq \frac{\alpha^{2}}{\epsilon}(\epsilon^{\frac{1}{2}}\|u_{x}(s)\|_{L^{2}})^{2}
\leq \alpha^{2}\|u(s)\|_{H^{1}}^{2}.
\end{array}
\eqno(4.22)
$$
Combining (4.21) and (4.22), we prove that the integral in (4.20) approaches zero
as $\epsilon\rightarrow 0$. We now estimate the integral near the corners of the domain
$$
\begin{array}{rl}
&|(\int_{\tau_{0}-\epsilon}^{\tau_{0}}+\int_{\tau}^{\tau+\epsilon})\int_{x(s)}^{x(s)+\epsilon}
u_{x}[\psi^{\epsilon}_{t}
+(\alpha u+\beta)\psi^{\epsilon}_{x}](s,z)dzds|\\[3pt]
&\leq (\int_{\tau_{0}-\epsilon}^{\tau_{0}}+\int_{\tau}^{\tau+\epsilon})
(\int_{x(s)}^{x(s)+\epsilon}u_{x}^{2}(s,z)dz)^{\frac{1}{2}}
(\int_{x(s)}^{x(s)+\epsilon}[\psi^{\epsilon}_{t}
+(\alpha u+\beta)\psi^{\epsilon}_{x}]^{2}(s,z)dz)^{\frac{1}{2}}ds\\[3pt]
&\leq 2\epsilon\|u(s)\|_{H^{1}}
(\int_{x(s)}^{x(s)+\epsilon}4\epsilon^{-2}(1+|\alpha|\|u(s)\|_{L^{\infty}}+|\beta|)
dx)^{\frac{1}{2}}
\leq C\epsilon^{\frac{1}{2}}\rightarrow 0
\end{array}
$$
as $\epsilon\rightarrow 0$. The above analysis shows that
$$
\lim_{\epsilon\rightarrow 0}\int_{\tau_{0}-\epsilon}^{\tau+\epsilon}\int_{x(s)}^{x(s)+\epsilon}
u_{x}[\psi^{\epsilon}_{t}
+(\alpha u+\beta)\psi^{\epsilon}_{x}](s,z)dzds=0.
$$
Thus, from (4.19) we know that (4.8) holds for all $0\leq \tau_{0}<\tau<\tilde{T}$. Since $\tilde{T}$ is arbitrary, (4.8) holds for all $0\leq \tau_{0}<\tau<\infty$.

Finally, we prove the uniqueness of $x(t)$. Assume that there exist
two different $x_{1}(t)$ and $x_{2}(t)$, which satisfy (4.6) and (4.7).
Now, choosing the measurable functions $\eta_{1}$ and $\eta_{2}$
such that $x_{1}(t)=x(t,\eta_{1}(t))$ and $x_{2}(t)=x(t,\eta_{2}(t))$.
Then, $\eta_{1}(\cdot)$ and $\eta_{2}(\cdot)$ satisfy (4.11) with the same
initial data $x(0)=x_{0}$. This contradicts with the uniqueness of $\eta$.
\hfill $\Box$\\

\noindent\textbf{Lemma 4.3.}
Let $u=u(t,x)$ be the weak solution of (1.7) satisfying (1.10)-(1.11), and $t\mapsto \eta(t;\tau,\bar{\eta})$ be the solution to the integral equation
$$
\begin{array}{rl}
\eta(t)
=\bar{\eta}+\int_{\tau}^{t}G(s,\eta(s))ds,
\end{array}
\eqno(4.23)
$$
where $G$ is defined in (4.10). Then the following results hold:

(i) the map $(t,\eta)\mapsto u(t,\eta):=u(t,x(t,\eta))$ is locally Lipschitz continuous with a constant depending on $\|u_{0}\|_{H^{1}}$ and the time interval;

(ii) for any two initial data $\bar{\eta}_{1}$
and $\bar{\eta}_{2}$, and any times $t,\tau\geq 0$, there exists a constant $C$
such that the corresponding solutions satisfy
$$
|\eta(t;\tau,\bar{\eta}_{1})-\eta(t;\tau,\bar{\eta}_{2})|
\leq e^{C|t-\tau|}|\bar{\eta}_{1}-\bar{\eta}_{2}|.
$$

\noindent\textbf{Proof.}
(i) For all $0\leq\tau<t<\tilde{T}$ with any fixed $\tilde{T}>0$, it follows from (4.5),(4.8) and (4.11) that
$$
\begin{array}{rl}
|u(t,x(t,\bar{\eta}))-u(\tau,\bar{\eta})|
&\leq |u(t,x(t,\bar{\eta}))-u(t,x(t,\eta(t)))|
+|u(t,x(t,\eta(t)))-u(\tau,x(\tau,\eta(\tau)))|\\[3pt]
&\leq \frac{1}{2}\max\{1,e^{-2\lambda t}\}|\eta(t)-\bar{\eta}|
+C(t-\tau)
\leq C(t-\tau),
\end{array}
$$
where $C$ is a constant depending on $\|u_{0}\|_{H^{1}}$ and $\tilde{T}$.

(ii) For all $0\leq\tau<t<\tilde{T}$ with any fixed $\tilde{T}>0$, it follows from the Lipschitz continuity of $G$ that
$$
\begin{array}{rl}
|\eta(t;\tau,\bar{\eta}_{1})-\eta(t;\tau,\bar{\eta}_{2})|
\leq |\bar{\eta}_{1}-\bar{\eta}_{2}|
+C\int_{\tau}^{t}|\eta(s;\tau,\bar{\eta}_{1})-\eta(s;\tau,\bar{\eta}_{2})|ds
\leq |\bar{\eta}_{1}-\bar{\eta}_{2}|e^{C(t-\tau)},
\end{array}
$$
where the last inequality is obtained by using Gronwall's lemma and $C$ is a constant depending on $\|u_{0}\|_{H^{1}}$ and $\tilde{T}$.  \hfill $\Box$\\

\noindent\textbf{Proof of Theorem 1.2.}
\emph{Step 1.} It follows from Lemmas 4.1 and 4.3 that the map
$(t,\eta)\mapsto (x,u)(t,\eta)$ is locally Lipschitz continuous.
According to an entirely similar argument we find that
the maps $\eta\mapsto G(t,\eta):=G(t,x(t,\eta))$
and $\eta\mapsto \partial_{x}P_{i}(t,\eta):=\partial_{x}P_{i}(t,x(t,\eta))$
are also Lipschitz continuous. By Rademacher's theorem, the partial derivatives
$x_{t}, x_{\eta}, u_{t}, u_{\eta}, G_{\eta}$ and $\partial_{\eta}(\partial_{x}P_{i})$
exist almost everywhere. Moreover, almost every point $(t,\eta)$ is a Lebesgue point
for these derivatives. Let $t\mapsto \eta(t,\bar{\eta})$ be the unique solution
to the integral equation (4.11), then from Lemma 4.3 we know the following statement holds for almost every $\bar{\eta}$: \\

\noindent\textbf{(GC)} For a.e. $t>0$, the point $(t, \eta(t,\bar{\eta}))$ is a Lebesgue point for the partial derivatives $x_{t}, x_{\eta}, u_{t}, u_{\eta}, G_{\eta}$
and $\partial_{\eta}(\partial_{x}P_{i})$. Moreover,
$x_{\eta}(t, \eta(t,\bar{\eta}))>0$ for a.e. $t>0$. \\

\noindent If (GC) holds, we then say that $t\mapsto \eta(t,\bar{\eta})$ is a \emph{good characteristic}.

\emph{Step 2.} We seek an ODE describing how the quantities $u_{\eta}$ and $x_{\eta}$ vary along
a good characteristic. As in Lemma 4.3, we denote by $t\mapsto \eta(t;\tau,\bar{\eta})$
the solution to (4.23). If $t,\tau\notin \mathcal{N}$,
assuming that $\eta(\cdot; \tau, \bar{\eta})$ is a good characteristic, differentiating (4.23) w.r.t. $\bar{\eta}$ we find
$$
\begin{array}{l}
\frac{\partial}{\partial \bar{\eta}}\eta(t; \tau, \bar{\eta})
=1+\int_{\tau}^{t}G_{\eta}(s,\eta(s;\tau,\bar{\eta}))\cdot
\frac{\partial}{\partial \bar{\eta}}\eta(s; \tau, \bar{\eta})ds.
\end{array}
\eqno(4.24)
$$
Next, differentiating the following identity w.r.t. $\bar{\eta}$
$$
x(t,\eta(t; \tau, \bar{\eta}))
=x(\tau, \bar{\eta})+\int_{\tau}^{t}[\alpha u(s,x(s,\eta(t; \tau, \bar{\eta})))+\beta]ds,
$$
we obtain
$$
\begin{array}{l}
x_{\eta}(t,\eta(t; \tau, \bar{\eta}))\cdot
\frac{\partial}{\partial \bar{\eta}}\eta(t; \tau, \bar{\eta})
=x_{\eta}(\tau, \bar{\eta})
+\int_{\tau}^{t}\alpha u_{\eta}(s,\eta(s; \tau, \bar{\eta}))\cdot
\frac{\partial}{\partial \bar{\eta}}\eta(s; \tau, \bar{\eta})ds.
\end{array}
\eqno(4.25)
$$
Since (4.8) holds, by using the notation $u(t,\eta):=u(t,x(t,\eta))$, we have
$$
\begin{array}{rl}
u(t,\eta(t; \tau_{0}, \bar{\eta}))-u(\tau_{0},\bar{\eta})
=-\int_{\tau_{0}}^{t}(\partial_{x}P_{1}+P_{2}+\lambda u)(s,\eta(s; \tau_{0}, \bar{\eta}))ds.
\end{array}
$$
Differentiating the above identity w.r.t. $\bar{\eta}$, we obtain
$$
\begin{array}{l}
u_{\eta}(t,\eta(t; \tau, \bar{\eta}))\cdot
\frac{\partial}{\partial \bar{\eta}}\eta(t; \tau, \bar{\eta})
=u_{\eta}(\tau, \bar{\eta})
-\int_{\tau}^{t}(\partial_{x}P_{1}+P_{2}+\lambda u)_{\eta}(s,\eta(s; \tau, \bar{\eta}))
\cdot
\frac{\partial}{\partial \bar{\eta}}\eta(s; \tau, \bar{\eta})ds.
\end{array}
\eqno(4.26)
$$
Combining (4.24)-(4.26), we thus obtain the system of ODEs
$$
\begin{array}{l}
\left\{\begin{array}{l}
\frac{d}{dt}\left[\frac{\partial}{\partial \bar{\eta}}\eta(t; \tau, \bar{\eta})\right]
=G_{\eta}(t,\eta(t;\tau,\bar{\eta}))\cdot
\frac{\partial}{\partial \bar{\eta}}\eta(t; \tau, \bar{\eta}),\\[3pt]
\frac{d}{dt}\left[x_{\eta}(t,\eta(t; \tau, \bar{\eta}))\cdot
\frac{\partial}{\partial \bar{\eta}}\eta(t; \tau, \bar{\eta})\right]
=\alpha u_{\eta}(t,\eta(t; \tau, \bar{\eta}))\cdot
\frac{\partial}{\partial \bar{\eta}}\eta(t; \tau, \bar{\eta}),\\[3pt]
\frac{d}{dt}\left[u_{\eta}(t,\eta(t; \tau, \bar{\eta}))\cdot
\frac{\partial}{\partial \bar{\eta}}\eta(t; \tau, \bar{\eta})\right]
=-(\partial_{x}P_{1}+P_{2}+\lambda u)_{\eta}(t,\eta(t; \tau, \bar{\eta}))
\cdot
\frac{\partial}{\partial \bar{\eta}}\eta(t; \tau, \bar{\eta}).
\end{array}
\right.
\end{array}
\eqno(4.27)
$$
In particular, the quantities within square brackets on the left hand sides
of (4.27) are absolutely continuous.
Recall that the fact $u_{x}^{2}=e^{-2\lambda t}\frac{1-x_{\eta}}{x_{\eta}}$
as long as $x_{\eta}>0$.
From (4.27), along a good characteristic
we obtain
$$
\begin{array}{l}
\left\{\begin{array}{l}
\frac{d}{dt}x_{\eta}+G_{\eta}x_{\eta}
=\alpha u_{\eta},\\[3pt]
\frac{d}{dt}u_{\eta}+(G_{\eta}+\lambda)u_{\eta}
=-(\partial_{x}P_{1}+P_{2})_{x}x_{\eta}
=(h(u)-P_{1}-\partial_{x}P_{2}-\frac{\alpha}{2}e^{-2\lambda t})x_{\eta}
+\frac{\alpha}{2}e^{-2\lambda t}.
\end{array}
\right.
\end{array}
$$

\emph{Step 3.} We revert to the original $(t,x)$ coordinates and deduce an evolution equation for the partial derivative $u_{x}$ along a good characteristic curve.

Fix a point $(\tau, \bar{x})$ with $\tau\notin \mathcal{N}$. Suppose that $\bar{x}$ is a Lebesgue point for the map $x\mapsto u_{x}(\tau,x)$. Let $\bar{\eta}$ be the coordinate value satisfying
$\bar{x}=x(\tau,\bar{\eta})$ and assume that $t\mapsto \eta(t;\tau,\bar{\eta})$ is
a good characteristic, so that (GC) holds. We observe that
$$
u_{x}^{2}(\tau, \bar{x})=e^{-2\lambda \tau}(\frac{1}{x_{\eta}(\tau,\bar{\eta})}-1)\geq 0,
$$
which implies $0<x_{\eta}(\tau,\bar{\eta})\leq1$.
If $t_{0}$ is any time where $x_{\eta}(t_{0},\eta(t_{0}; \tau, \bar{\eta}))>0$, we can find a neighborhood $I=[t_{0}-\delta, t_{0}+\delta]$ such that $x_{\eta}(t,\eta(t; \tau, \bar{\eta}))>0$ on $I$.
Using the two ODEs (4.25)-(4.26) describing the evolution of $u_{\eta}$ and $x_{\eta}$, we conclude that the map
$$
t\mapsto u_{x}(t,\eta(t; \tau, \bar{\eta}))
:=\frac{u_{\eta}(t, \eta(t; \tau, \bar{\eta}))}{x_{\eta}(t, \eta(t; \tau, \bar{\eta}))}
$$
is absolutely continuous on $I$ and satisfies
$$
\begin{array}{rl}
&\dfrac{d}{dt}u_{x}(t,\eta(t; \tau, \bar{\eta}))
=\dfrac{d}{dt}(\dfrac{u_{\eta}}{x_{\eta}})
=\frac{x_{\eta}[(h(u)-P_{1}-\partial_{x}P_{2}-\frac{\alpha}{2}e^{-2\lambda t})x_{\eta}
+\frac{\alpha}{2}e^{-2\lambda t}-(G_{\eta}+\lambda)u_{\eta}]
-u_{\eta}(\alpha u_{\eta}-G_{\eta}x_{\eta})}{x_{\eta}^{2}}\\[3pt]
&=h(u)-P_{1}-\partial_{x}P_{2}-\frac{\alpha}{2}e^{-2\lambda t}
+\frac{\alpha}{2}e^{-2\lambda t}\frac{1}{x_{\eta}}
-\frac{G_{\eta}u_{\eta}}{x_{\eta}}-\lambda\frac{u_{\eta}}{x_{\eta}}
-\frac{\alpha u_{\eta}^{2}}{x_{\eta}^{2}}
+\frac{G_{\eta}u_{\eta}}{x_{\eta}}\\[3pt]
&=h(u)-P_{1}-\partial_{x}P_{2}-\frac{\alpha}{2}e^{-2\lambda t}
+\frac{\alpha}{2}e^{-2\lambda t}\frac{1}{x_{\eta}}
-\lambda\frac{u_{\eta}}{x_{\eta}}
-\frac{\alpha u_{\eta}^{2}}{x_{\eta}^{2}}\\[3pt]
&=h(u)-P_{1}-\partial_{x}P_{2}
+\frac{\alpha}{2}e^{-2\lambda t}
-\frac{\alpha}{2}e^{-2\lambda t}\frac{1}{x_{\eta}}
-\lambda u_{x}.
\end{array}
$$
In turn, as long as $x_{\eta}>0$ this implies
$$
\begin{array}{rl}
&\dfrac{d}{dt}\arctan u_{x}(t,\eta(t; \tau, \bar{\eta}))
=\dfrac{1}{1+u_{x}^{2}}\dfrac{d}{dt}u_{x}(t,\eta(t; \tau, \bar{\eta}))\\[3pt]
&=\left(h(u)-P_{1}-\partial_{x}P_{2}
+\frac{\alpha}{2}e^{-2\lambda t}
-\frac{\alpha}{2}e^{-2\lambda t}\frac{1}{x_{\eta}}
-\lambda u_{x}\right)\dfrac{1}{1+e^{-2\lambda t}(\frac{1}{x_{\eta}}-1)}\\[3pt]
&=\left(h(u)-P_{1}-\partial_{x}P_{2}+\frac{\alpha}{2}-\lambda u_{x}\right)\dfrac{1}{1+e^{-2\lambda t}(\frac{1}{x_{\eta}}-1)}-\frac{\alpha}{2}.
\end{array}
\eqno(4.28)
$$

Now we consider the function
$$
\begin{array}{rl}
v:=
\left\{\begin{array}{l}
2\arctan u_{x}, \quad \mbox{if}~0<x_{\eta}\leq 1, \\[3pt]
\pi, \quad \mbox{if}~x_{\eta}=0.
\end{array}
\right.
\end{array}
$$
This implies
$$
\begin{array}{rl}
\dfrac{x_{\eta}}{x_{\eta}+e^{-2\lambda t}(1-x_{\eta})}
=\dfrac{1}{1+u_{x}^{2}}=\cos^{2}\frac{v}{2}, \quad
\dfrac{u_{x}}{1+u_{x}^{2}}=\frac{1}{2}\sin v, \quad
\dfrac{u_{x}^{2}}{1+u_{x}^{2}}=\sin^{2}\frac{v}{2}.
\end{array}
\eqno(4.29)
$$
Then, $v$ will be regarded as map taking values in the unit circle
$\Omega:=[-\pi, \pi]$ with endpoints identified. We claim that, along
each good characteristic, the map $t\mapsto v(t):=v(t,x(t,\eta(t;\tau,\bar{\eta})))$
is absolutely continuous and satisfies
$$
\begin{array}{rl}
\frac{d}{dt}v(t)
=2\left(h(u)-P_{1}-\partial_{x}P_{2}+\frac{\alpha}{2}\right)\cos^{2}\frac{v}{2}
-\lambda \sin v-\alpha.
\end{array}
\eqno(4.30)
$$
Indeed, denote by $x_{\eta}(t)$, $u_{\eta}(t)$ and $u_{x}(t)=\frac{u_{\eta}(t)}{x_{\eta}(t)}$ the values of $x_{\eta}$,
$u_{\eta}$ and $u_{x}$ along this particular characteristic. By (GC) we have $x_{\eta}>0$ for a.e. $t>0$. If $t_{0}$ is any time where $x_{\eta}(t_{0})>0$, we can find a neighborhood $I=[t_{0}-\delta,t_{0}+\delta]$ such that $x_{\eta}(t)>0$ on $I$. By (4.28) and (4.29), $v=2\arctan(\frac{u_{\eta}}{x_{\eta}})$ is absolutely continuous restricted to
$I$ and satisfies (4.30). To prove our claim, it thus remains to show that $t\mapsto v(t)$ is continuous on the null set $\mathcal{N}$ of times where $x_{\eta}(t)=0$.

Suppose $x_{\eta}(t_{0})=0$. From the fact that the identity
$u_{x}^{2}=e^{-2\lambda t}\frac{1-x_{\eta}}{x_{\eta}}$ holds as long as $x_{\eta}>0$,
it is clear that $x_{\eta}(t)\rightarrow 0$ and $u_{x}^{2}(t)\rightarrow +\infty$ as $t\rightarrow t_{0}$. This implies $v(t)=2\arctan u_{x}(t)\rightarrow \pm\pi$.
Since we identity the points $\pm\pi$, we know $v$ is continuous at $t_{0}$,
proving our claim.

\emph{Step 4.} Now let $u=u(t,x)$ be a global weak solution of (1.7) satisfying (1.10)-(1.11). As shown by the previous analysis, in terms of the variables $t, \eta$ the quantities $x,u,v$ satisfy
the semi-linear system
$$
\begin{array}{l}
\left\{\begin{array}{l}
\frac{d}{dt}\eta(t,\bar{\eta})
=G(t,\eta(t,\bar{\eta})),\\[3pt]
\frac{d}{dt}x(t,\eta(t,\bar{\eta}))
=\alpha u(t,\eta(t,\bar{\eta}))+\beta,\\[3pt]
\frac{d}{dt}u(t,\eta(t,\bar{\eta}))
=-(\partial_{x}P_{1}+P_{2}+\lambda u)(t,\eta(t,\bar{\eta})),\\[3pt]
\frac{d}{dt}v(t,\eta(t,\bar{\eta}))
=2\left(h(u)-P_{1}-\partial_{x}P_{2}+\frac{\alpha}{2}\right)\cos^{2}\frac{v}{2}
-\lambda \sin v-\alpha,
\end{array}
\right.
\end{array}
\eqno(4.31)
$$
where $P_{1},P_{2},\partial_{x}P_{1}$ and $\partial_{x}P_{2}$
admit representations in terms of the variable $\eta$, namely
$$
\begin{array}{rl}
P_{1}(x(\eta))
&=\frac{1}{2}\int_{-\infty}^{+\infty}
e^{-\left|\int_{\eta}^{\eta^{\prime}}\frac{\cos^{2}\frac{v}{2}}
{\cos^{2}\frac{v}{2}+e^{2\lambda t}\sin^{2}\frac{v}{2}}(t,\hat{\eta})d\hat{\eta}\right|}\cdot\\[3pt]
&\qquad\left[\frac{1}{\cos^{2}\frac{v}{2}+e^{2\lambda t}\sin^{2}\frac{v}{2}}
(h(u)\cos^{2}\frac{v}{2}
+\frac{\alpha}{2}\sin^{2}\frac{v}{2})\right](t,\eta^{\prime})d\eta^{\prime},
\end{array}
$$

$$
\begin{array}{rl}
\partial_{x}P_{1}(x(\eta))
&=\frac{1}{2}(\int_{\eta}^{+\infty}-\int_{-\infty}^{\eta})
e^{-\left|\int_{\eta}^{\eta^{\prime}}\frac{\cos^{2}\frac{v}{2}}
{\cos^{2}\frac{v}{2}+e^{2\lambda t}\sin^{2}\frac{v}{2}}(t,\hat{\eta})d\hat{\eta}\right|}\cdot\\[3pt]
&\qquad\left[\frac{1}{\cos^{2}\frac{v}{2}+e^{2\lambda t}\sin^{2}\frac{v}{2}}
(h(u)\cos^{2}\frac{v}{2}
+\frac{\alpha}{2}\sin^{2}\frac{v}{2})\right](t,\eta^{\prime})d\eta^{\prime},
\end{array}
$$

$$
\begin{array}{rl}
P_{2}(x(\eta))
&=\frac{k}{2}\int_{-\infty}^{+\infty}
e^{-\left|\int_{\eta}^{\eta^{\prime}}\frac{\cos^{2}\frac{v}{2}}
{\cos^{2}\frac{v}{2}+e^{2\lambda t}\sin^{2}\frac{v}{2}}(t,\hat{\eta})d\hat{\eta}\right|}\cdot\\[3pt]
&\qquad\left(\frac{1}{\cos^{2}\frac{v}{2}+e^{2\lambda t}\sin^{2}\frac{v}{2}}u\cos^{2}\frac{v}{2}\right)(t,\eta^{\prime})d\eta^{\prime},
\end{array}
$$

$$
\begin{array}{rl}
\partial_{x}P_{2}(x(\eta))
&=\frac{k}{2}(\int_{\eta}^{+\infty}-\int_{-\infty}^{\eta})
e^{-\left|\int_{\eta}^{\eta^{\prime}}\frac{\cos^{2}\frac{v}{2}}
{\cos^{2}\frac{v}{2}+e^{2\lambda t}\sin^{2}\frac{v}{2}}(t,\hat{\eta})d\hat{\eta}\right|}\cdot\\[3pt]
&\qquad\left(\frac{1}{\cos^{2}\frac{v}{2}+e^{2\lambda t}\sin^{2}\frac{v}{2}}u\cos^{2}\frac{v}{2}\right)(t,\eta^{\prime})d\eta^{\prime}.
\end{array}
$$
For every $\bar{\eta}\in \mathbb{R}$ we have the initial condition
$$
\begin{array}{l}
\left\{\begin{array}{l}
\eta(0,\bar{\eta})
=\bar{\eta},\\[3pt]
x(0,\bar{\eta})
=x(0,\bar{\eta}),\\[3pt]
u(0,\bar{\eta})
=u_{0}(x(0,\bar{\eta})),\\[3pt]
v(0,\bar{\eta})
=2\arctan u_{0,x}(x(0,\bar{\eta})).
\end{array}
\right.
\end{array}
\eqno(4.32)
$$
By the previous lemmas, it is easy to prove the Lipschitz continuity of all coefficients. Consequently, the Cauchy problem
(4.31)-(4.32) has a unique global solution defined for all $t>0$
and $x\in \mathbb{R}$.

\emph{Step 5.} To finish the proof of the uniqueness, suppose that there exist
two weak solutions $u$ and $\hat{u}$ to the Cauchy problem (1.7)
with the same initial data $u_{0}\in H^{1}(\mathbb{R})$.
We know that the related Lipschitz continuous maps
$\eta\mapsto x(t,\eta)$ and $\eta\mapsto \hat{x}(t,\eta)$ are
strictly increasing for a.e. $t\geq 0$. Thus they have continuous
inverses $x\mapsto \eta^{*}(t,x), x\mapsto \hat{\eta}^{*}(t,x)$.
By performing the previous analysis, the map $(t,\eta)\mapsto (x,u,v)(t,\eta)$
is uniquely determined by the initial data $u_{0}$.
Therefore
$$
x(t,\eta)=\hat{x}(t,\eta),\quad
u(t,\eta)=\hat{u}(t,\eta),
$$
which implies that, for a.e. $t\geq 0$,
$$
u(t,x)=u(t,\eta^{*}(t,x))
=\hat{u}(t,\hat{\eta}^{*}(t,x))
=\hat{u}(t,x).
$$
This completes the proof of Theorem 1.2.  \hfill $\Box$

\section*{Acknowledgments}
Chen's work was supported by NSFC (No:11801432). Li's work was supported by NSFC (No:11571057).
Wang's work was supported by NSFC (No:11801429) and the Natural Science Basic Research Plan in Shaanxi Province of China (No:2019JQ-136).

\label{}





\bibliographystyle{model3-num-names}
\bibliography{<your-bib-database>}



\end{document}